\documentclass[a4paper, 11pt]{article}
\usepackage[titletoc,toc,title]{appendix}
\usepackage{fullpage} 

\usepackage{morefloats} 
\usepackage{graphicx}
\usepackage{dsfont}
\usepackage{color}
\usepackage{xcolor}
\usepackage{mathtools}
\usepackage{upgreek}
\usepackage{longtable}
\usepackage[justification=justified,
            format=plain]{caption}
\usepackage{natbib}
\allowdisplaybreaks

\usepackage{mathtools}
\usepackage{enumitem}

\usepackage{lineno}

\newcommand{\dd}[1]{\frac{\mathrm{d}}{\mathrm{d}#1}}

\newcommand{\rpar}[1]{\left(#1\right)}

\newcommand{\functionof}[2]{#1\left(#2\right)}

\newcommand{\expp}[1]{\mathrm{e}^{#1}}
\DeclareMathOperator\erf{erf}

\newcommand{\Hill}[2]{\ensuremath{\mathcal{H}\left(#1,#2\right)}}

\newcommand*{\figref}[1]{\figurename~\ref{#1}}
\newcommand*{\secref}[1]{Section~\ref{#1}}
\newcommand*{\appref}[1]{Appendix~\ref{#1}}
\renewcommand*{\eqref}[1]{equation~\ref{#1}}

\title{Modulation of synaptic plasticity by glutamatergic gliotransmission: A modeling study}
\author{
        Maurizio De~Pitt\`a \\
                Department of Neurobiology\\
        The University of Chicago, Chicago, IL 60637, USA\\
                Project-Team BEAGLE, INRIA Rh\^{o}ne-Alpes, 60097 Villeurbanne, France
            \and
        Nicolas Brunel\\
                Departments of Statistics and Neurobiology\\
        The University of Chicago, Chicago, IL 60637, USA\\
        }
\date{\today}

\begin{document}
\maketitle

\begin{abstract}
Glutamatergic gliotransmission, that is the release of glutamate from
perisynaptic astrocyte processes in an activity-dependent manner, has
emerged as a potentially crucial signaling pathway for regulation of
synaptic plasticity, yet its modes of expression and function in vivo
remain unclear. Here, we focus on two experimentally well-identified
gliotransmitter patwhays: (i)~modulations of synaptic release and
(ii)~postynaptic slow inward currents mediated by glutamate released
from astrocytes, and investigate their possible functional relevance
on synaptic plasticity in a biophysical model of an
astrocyte-regulated synapse. Our model predicts that both pathways
could profoundly affect both short- and long-term plasticity. In
particular, activity-dependent glutamate release from astrocytes,
could dramatically change spike-timing--dependent plasticity, turning
potentiation into depression (and vice versa) for the same protocol.
\end{abstract}

\newpage
\section*{Abbreviations}
AMPAR:~$\alpha$-amino-3-hydroxy-5-methyl-4-isoxazolepropionic acid receptor; AP:~action potential; bAP:~back-propagating action potential; ER:~endoplasmic reticulum; GPCR:~G~protein-coupled receptor; LTD:~long-term depression; LTP:~long-term potentiation; mGluR:~metabotropic glutamate receptor; NMDA(R):~N-methyl-\textsc{d}-aspartate (receptors); PAR1:~protease-activated receptor~1; PPR:~pair pulse ratio (E)PSC:~(excitatory) postsynaptic current; (E)PSP:~(excitatory) postsynaptic potential; SIC:~slow inward current; SERCA:~sarco-endoplasmic recticulum Ca$^{2+}$/ATPase; STDP:~spike-timing--dependent plasticity; VDCC:~voltage-dependent calcium channel.

\newpage
\section*{Introduction}
In recent years, astrocytes have attracted great interest for their
capacity to release neuroactive molecules, among which are
neurotransmitters like glutamate, because these molecules could
modulate neural activity and lead to a possible role for astrocytes in
neural information processing
\citep{VolterraMeldolesiRev2005,PereaAraque_Science2007,HalassaHaydon_ARP2010}. Indeed,
astrocyte-derived neurotransmitters, also called ``gliotransmitters''
for their astrocytic origin \citep{BezziVolterraRev2001}, have been
shown to act on neurons and to regulate synaptic transmission and
plasticity through a variety of mechanisms
\citep{Araque_Neuron2014}. The binding of receptors located on either
pre- or postsynaptic terminals by astrocyte-released glutamate has
historically been the first pathway for gliotransmission to be
discovered and, arguably, the most studied one experimentally for its
several possible functional implications
\citep{SantelloVolterra_Neurosci2009}.

Activation of extrasynaptic receptors on presynaptic terminals by
astrocytic glutamate modulates the probability of neurotransmitter
release from those terminals \citep{SantelloVolterra_Neurosci2009}. In
particular, depending on receptor type, such modulation may be either
toward an increase or toward a decrease of the frequency of
spontaneous
\citep{FiaccoMcCarthy2004,JourdainVolterra2007,Bonansco_etal_EJN2011,Panatier_etal_Cell2011,Perea_NatComm2014}
and evoked neurotransmitter release both in excitatory
\citep{JourdainVolterra2007,PereaAraque_Science2007,Navarrete_Neuron2010,Panatier_etal_Cell2011}
and inhibitory synapses
\citep{LiuNedergaard2004,Liu_PNAS2004,Benedetti_JP2011astrocytes}. Because
synaptic release probability characterizes how a synapse filters or,
in other words, ``processes'' presynaptic action potentials
\citep{MarkramWangTsodyks_PNAS1998,AbbottRegehr_Nat2004}, modulations
of synaptic release probability by astrocytic glutamate are suggested
to alter the computational properties of neural circuits
\citep{DePitta_Neurosci2015}.

Glutamate released by astrocytes may also bind to
extrasynaptically-located postsynaptic NMDA receptors, evoking slow
inward currents~(SICs) in nearby neurons
\citep{ParriCrunelli2001,Angulo2004,FellinCarmignoto_Neuron2004,PereaAraque_JN2005,DAscenzo_PNAS2007,Shigetomi_JN2008,Bardoni_JP2010,Perea_NatComm2014,Martin_Science2015}. The
depolarizing action of these currents modulates neural excitability
with the potential to affect neuronal action potential firing
\citep{Halassa_TMM2007}. Moreover, because single astrocytes are in
close proximity to a large number ($\sim$100) of neurons
\citep{HalassaHaydon2007}, it has been suggested that an inward
current can be generated in many adjacent neurons, thereby promoting
synchrony of neuronal firing
\citep{ParriCrunelli2001,Angulo2004,FellinCarmignoto_Neuron2004}.

Although modulations of both synaptic release and~SICs mediated by
glutamatergic gliotransmission have been recorded in the cortex and
the hippocampus, as well as in several other brain regions
\citep{Araque_Neuron2014}, their physiological relevance remains
elusive. In particular, beyond regulation of synaptic filtering and
neuronal firing, theoretical arguments support a further possible role
for both pathways in the regulation of NMDAR-mediated
spike-timing--dependent plasticity (STDP) \citep{DePitta_FCN13}. Both
pathways clearly have the potential to regulate activation of
postsynaptic NMDA receptors: the former does so indirectly, by
modulations of the amount of synaptically-released neurotransmitter
molecules that bind to~NMDA receptors in the synaptic cleft; the
latter directly, by targeting extrasynaptic NMDA receptors. Thus, by
controlling postsynaptic NMDAR activation, glutamatergic
gliotransmission could ultimately regulate the~STDP outcome, that is
either potentiation (LTP) or depression (LTD)
\citep{Mizuno_JN2001,Nevian_JN2006}. Consistent with this hypothesis,
experiments have reported a lower threshold for~LTP induction at
hippocampal synapses when synaptic release is increased by astrocytic
glutamate \citep{Bonansco_etal_EJN2011}. And long-term potentiation of
orientation-selective responses of neurons in the primary visual
cortex by cholinergic activation of surrounding astrocytes, has also
been reported to be correlated with an increase of~SIC frequency in
those neurons \citep{Chen_PNAS2012}.

While the potential impact on~STDP of pre- or postsynaptic
activity-dependent modulations of synaptic efficacy have widely been
addressed both experimentally \citep{Sjostrom_PR2008} and
theoretically \citep{Froemke_FSN2010,Graupner_FCN2010}, the possible
effect on plasticity of the regulation of these modulations by
glutamatergic gliotransmission (and by gliotransmission in general)
has been investigated by very few theoretical studies. These studies
suggest a potential role in~LTP induction both for large increases of
synaptic release and for large~SICs mediated by astrocytic glutamate
\citep{Wade_PLoSOne2011,Naeem_TNNLS2015}. This scenario seems however
at odds with the majority of recent experimental observations that
report modest signaling magnitudes for these two routes of
gliotransmission. It is thus not clear under what biophysical
conditions, modulations of synaptic release or~SICs mediated by
glutamatergic gliotransmission could
affect~STDP. Astrocyte-mediated~SICs, for example, are known to occur
sporadically, being recorded in single neurons only as often as
$<$5/min \citep{Chen_PNAS2012,Martin_Science2015}, raising the
question whether and how, by occurring at such low rates, they could
effectively play a role in~STDP.

We thus set to investigate what conditions are required for
glutamatergic gliotransmission to affect~STDP by presynaptic
modulations of neurotransmitter release or through
postsynaptic~SICs. We extend the model of an astrocyte-regulated
synapse originally introduced by \citet{DePitta_PCB11} to include a
biophysically-realistic description of synaptically-evoked
gliotransmitter release by the astrocyte as well as a mechanism for
the generation of postsynaptic~SICs and~STDP. Extensive numerical
investigations of our model leads to two major predictions. First,
glutamatergic gliotransmission could change the nature of~STDP by
modifying the parameter ranges for~LTP and~LTD induction. Second, this
effect crucially depends on the nature of gliotransmission,
i.e. whether it is release-increasing vs. release-decreasing, its
strength, as well as its rate of occurrence and when it occurs with
respect to pre/post pairs. Thus, while glutamatergic gliotransmission
could potentially play a role in~STDP and learning, in practice this
effect must satisfy several biophysical and activity-dependent
constraints, supporting the existence of specialized dynamic
interactions between astrocytes and neurons.

\section*{Biophysical modelling of a gliotransmitter-regulated synapse}

Although there may be several possible routes by which astrocytes
release glutamate
\citep{NiParpura2007,ParpuraZorec_BRR2010,Zorec_ASN2012},~Ca$^{2+}$-dependent
glutamate release is likely the main one in physiological conditions
\citep{Barres_Neuron2008,Parpura_BBA2011}. From a modeling
perspective, as illustrated in
\figref{fig:signalling-pathway},~Ca$^{2+}$-dependent glutamatergic
gliotransmission consists of three distinct signaling pathways. One
pathway (\textit{black arrows}) initiates the
release-triggering~Ca$^{2+}$ signal in the astrocyte, and may be
either exogenous or heterosynaptic, or be triggered by the very synapses
that are modulated by glutamatergic gliotransmission in a homosynaptic
fashion. The other two pathways are instead represented by the two
recognized routes for the action of glutamatergic gliotransmission on
synaptic terminals: the presynaptic pathway whereby astrocytic
glutamate modulates synaptic release (\textit{magenta arrows}), and
the postsynaptic pathway which mediates~SICs in nearby neurons
(\textit{orange arrows}). Although both pathways could coexist at the same synapse in principle \citep{Perea_NatComm2014}, their
functional regulation is probably through
different~Ca$^{2+}$-dependent pathways \citep{Martin_Science2015},
both in terms of spatiotemporal~Ca$^{2+}$-dynamics
\citep{Shigetomi_JN2008} and in terms of pools of releasable glutamate
resources and/or mechanism of release for these latter
\citep{Hamilton_NRN2010}. Thus, in the following, we set to
investigate the effect of synaptic transmission of each pathway
independently of the other.

\subsection*{Calcium-dependent gliotransmitter release}

We begin our study by a description of a biophysically realistic model
of synaptically-evoked~Ca$^{2+}$-dependent glutamate release from an
astrocyte. At excitatory \citep{Perea_etal_TiNS2009} and inhibitory
synapses \citep{Losi_PTRSB2014}, astrocytes can respond to
synaptically-released neurotransmitters, by intracellular~Ca$^{2+}$
elevations and release glutamate in turn \citep{SantelloVolterra_Neurosci2009}. Although morphological and
functional details of the coupling between synaptic terminals and the
surrounding astrocytic processes remain to be fully elucidated, the
current hypothesis is that synaptically-evoked glutamate-releasing
astrocytic~Ca$^{2+}$ signaling is mainly by spillover of synaptic
neurotransmitters and/or other factors, which bind to high-affinity
astrocytic G~protein-coupled receptors (GPCRs)
\citep{Araque_Neuron2014} and thereby trigger inositol
1,4,5‐-trisphosphate (IP$_3$) production and~Ca$^{2+}$ release from the
endoplasmic reticulum (ER)
\citep{Nimmerjahn_JP2009,Volterra_NRN2014,Bazargani_NN2016}. While early work mainly
monitored somatic~Ca$^{2+}$ increases concluding that astrocytes
respond only to intense neuronal firing patterns \citep{Haydon2001},
recent experiments in astrocytic processes revealed that astrocytes
may also respond to low levels of synaptic activity by~Ca$^{2+}$
elevations confined in subcellular regions of their processes
\citep{DiCastro_Volterra_NatNeurosci2011,Panatier_etal_Cell2011,Bazargani_NN2016},
suggesting that the profile of astrocytic~Ca$^{2+}$ signaling, and
thus glutamate release that this latter could cause, encompass
the whole spectrum of neuronal (synaptic) activity
\citep{Araque_Neuron2014}.

To realistically describe synaptic release in the whole spectrum of
neuronal firing, we consider the model of an activity-dependent
synapse first introduced by \cite{TsodyksMarkram_PNAS1997}. This model
captures the dependence of synaptic release on past activity -- that
is presynaptic short-term plasticity -- which substantially influences
synaptic transmission at high enough rates of neuronal firing
\citep{ZuckerRegehrRev2002}. Accordingly, synaptic release results
from the product of the probability of having
neurotransmitter-containing vesicles available for release times the
probability of such vesicles to be effectively released by an action
potential \citep{DelCastilloKatz1954}, which correlates with
intrasynaptic~Ca$^{2+}$ \citep{SudhofRev2004}. At rest, it is assumed
that all vesicles are available for release. The arrival of an action
potential opens presynaptic voltage-dependent~Ca$^{2+}$
channels that trigger a transient increase of intrasynaptic~Ca$^{2+}$
which promotes release of a fraction~$u_S$ of available
vesicles. Following release, the emptied vesicles are refilled in
some characteristic time $\tau_d$, while
intrasynaptic~Ca$^{2+}$, and thus vesicle release probability, decay to
zero with a different time constant $\tau_f$. For multiple action
potentials incoming at time intervals of the order of these two time
constants, neither vesicle replenishment nor intrasynaptic~Ca$^{2+}$
are restored to their resting values, so that the resulting synaptic
release depends on the history of synaptic activity
\citep{Tsodyks_LesHouches2005}.

We illustrate the response of
the synapse model to a train of action potentials in Figures~\ref{fig:tripartite_io}A--C. The low rate of
stimulation of the first four action potentials
(\figref{fig:tripartite_io}A) allows for the reintegration of most of
the released neurotransmitter in between action potentials thereby keeping
vesicle depletion limited (\figref{fig:tripartite_io}B, \textit{orange trace}). In parallel, intrasynaptic~Ca$^{2+}$ grows, and so does vesicle release probability
(\figref{fig:tripartite_io}B, \textit{blue trace}), resulting in
progressively larger release of neurotransmitter per action potential or, in other words, in short-term facilitation of synaptic release
(\figref{fig:tripartite_io}C, $t<500$~ms). On the contrary, the
presentation of a series of action potentials in rapid succession at
$t=500$~ms, results in a sharp increase of vesicle release probability
to a value close to saturation (i.e.~Nt.~Rel.~Pr.$\simeq 1$) which causes
exhaustion of neurotransmitter resources (i.e.~Avail.~Nt.~Pr.$\simeq 0$). In this
scenario therefore, from one spike to the next one, progressively less
neurotransmitter is available for release and the amount of released
resources decreases with incoming action potentials, leading to
depression of synaptic transmission. Such depression is short-lived, since synaptic
release tends to recover after a sufficiently long period in which no
action potentials occur, that is the case, for example, of the last
action potential at $t=800$~ms.

Once released into the synaptic cleft, synaptic neurotransmitter is
rapidly cleared by diffusion as well as by other mechanisms, including uptake by transporters
and/or enzymatic degradation \citep{Clements1996,Diamond_JN2005}. In
the simplest approximation, the contribution of these mechanisms can
be modeled by a first order reaction \citep{DestexheJCompNeurosci1994}
which accounts for the exponentially decaying profile of
neurotransmitter concentration in \figref{fig:tripartite_io}C
after synaptic release at each action potential. A fraction of
released neurotransmitter molecules also spills out of the synaptic
cleft to the perisynaptic space (\figref{fig:tripartite_io}D) where it
binds to GPCRs on the astrocyte (\figref{fig:tripartite_io}E), therein
triggering~Ca$^{2+}$ signaling (\figref{fig:tripartite_io}F). To quantitatively describe this process, we modify the model of GPCR-mediated~Ca$^{2+}$ signaling originally introduced by
\citet{DePitta_JOBP2009} to account for dynamic
regulation of astrocytic receptors by synaptic activity (see
Appendix~\ref{app:model}, \secref{app-sec:ca-model}). Accordingly, as
illustrated in \figref{fig:tripartite_io}F, GPCR-mediated~Ca$^{2+}$
signaling is a result of the nonlinear interplay of three processes:
(i)~IP$_3$ production by GPCRs bound
by synaptic neurotransmitter (\textit{magenta trace}), (ii)~Ca$^{2+}$
release from the ER into the cytosol, which is triggered
by~IP$_3$-bound Ca$^{2+}$~channels (IP$_3$Rs) and also modulates
cytosolic~IP$_3$ (\textit{black trace}); and~(iii)~the effective
fraction of available, or more exactly, ``deinactivated''~IP$_3$Rs
\citep{DeYoungKeizerPNAS1992} that can take part in~Ca$^{2+}$ release
from the ER (\textit{yellow trace}). Depending on the choice of
parameter values, the astrocyte model may display both large,
long-lasting somatic~Ca$^{2+}$ elevations, and smaller and
shorter~Ca$^{2+}$ increases, akin to those reported in astrocytic
processes \citep{Volterra_NRN2014} (see
Appendix~\ref{app:par-estimation}).

Glutamate release from the astrocyte is then assumed to occur every
time that~Ca$^{2+}$ increases beyond a threshold concentration
(\figref{fig:tripartite_io}G, \textit{cyan dotted line}), in agreement
with experimental observations
\citep{Pasti1997,Marchaland_etal_JN2008}. Although different
mechanisms for glutamate release by the astrocyte could be possible, a
large amount of evidence points to vesicular exocytosis as the main
one to likely occur on a physiological basis
\citep{Sahlender_PTRSB2014}. Because astrocytic glutamate exocytosis
bears several similarities with its synaptic homologous (reviewed in
\citet{DePitta_FCN13}), we model it in the same fashion. Thus, in line
with experimental observations
\citep{BezziVolterra2004,BergersenGundersen_Neurosci2009}, we postulate
the existence of an astrocytic vesicular compartment that is
competent for regulated glutamate exocytosis. Then, upon a
suprathreshold~Ca$^{2+}$ elevation, a fixed fraction of astrocytic 
glutamate-containing vesicles is released into the extracellular
space and following reintegrated into the astrocyte with some
characteristic time constant (\figref{fig:tripartite_io}H). In this
fashion, glutamate concentration in the extracellular
space abruptly increases by exocytosis from the astrocyte, and then
exponentially decays akin to
neurotransmitter concentration in the synaptic cleft, yet, in general,
at a different rate (\figref{fig:tripartite_io}H) (Appendix~\ref{app:par-estimation}).

The description of gliotransmitter release hitherto introduced ignores
the possible stochastic nature of astrocytic glutamate release
\citep{Santello_etal_Neuron2011}, and reproduces the total amount of glutamate
released, on \emph{average}, by a single~Ca$^{2+}$ elevation beyond
the release threshold. This description provides a simplified general
framework to realistically capture synaptically-evoked glutamate
release by the astrocyte independently of the underlying mechanism of
astrocytic exocytosis, which may either be in the form of a burst of
synchronous vesicle fusion events that peaks within the first
50--500~ms from the~Ca$^{2+}$ rise underneath the plasma membrane
\citep{Domercq_etal_JBC2006,Marchaland_etal_JN2008,Santello_etal_Neuron2011},
or occur at slower fusion rates in an asynchronous fashion
\citep{Kreft2004,MalarkeyParpura_JP2011}.

\subsection*{Gliotransmitter-mediated regulation of synaptic release and short-term synaptic plasticity}

Once released, astrocyte-derived glutamate can diffuse in the
extracellular space and bind extrasynaptic receptors located on 
presynaptic terminals. In particular, ultrastructural evidence suggest
co-localization of glutamate-containing vesicles in perisynaptic
astrocytic processes with those receptors
\citep{JourdainVolterra2007}, hinting a focal action of astrocytic
glutamate on these latter. Such action is likely spatially confined
and temporally precise, akin to that of a neurotransmitter on postsynaptic receptors, and is not affected by synaptic neurotransmitters \citep{SantelloVolterra_Neurosci2009}. Both ionotropic and
metabotropic presynaptic receptors may be activated by astrocytic
glutamate, yet their differential recruitment likely depends on
developmental, regional, physiological and cellular (synaptic) factors
(reviewed in \citep{DePitta_FCN13}). The details of the biochemical
mechanisms of action of these receptors on synaptic physiology are not
fully understood \citep{PinheiroMulle_NatRev2008}, but the simplest
explanation is that they all modulate intrasynaptic~Ca$^{2+}$
levels eventually increasing or decreasing synaptic release probability
\citep{DePitta_Neurosci2015}, although in a receptor-specific fashion
\citep{ZuckerRegehrRev2002,PinheiroMulle_NatRev2008,Banerjee_TiNS2015}. 

From a modeling perspective, as originally proposed by
\citet{DePitta_PCB11}, the common effect on synaptic release shared by
different receptors allows to express, in the simplest approximation,
the synapse's resting release probability proportionally to the
fraction of presynaptic receptors activated by astrocytic glutamate
(\appref{app:model}, \secref{app-sec:ca-model}). In this fashion, as
illustrated in \figref{fig:pre-loop}, the time evolution of the
fraction of activated presynaptic receptors ensuing from a series of
glutamate release events by the astrocyte
(Figures~\ref{fig:pre-loop}A,B), is reflected by the dynamics of
synaptic release probability at rest averaged across different trials
(Figures~\ref{fig:pre-loop}C,E). The value of the coefficient of
proportionality for the dependence of synaptic release probability on
receptor activation sets the type of modulation of synaptic release by
astrocytic glutamate which can be either release-decreasing
(\figref{fig:pre-loop}C), such as in the case of astrocytic glutamate
binding presynaptic kainate receptors or group~II/III metabotropic
receptors (mGluRs)
\citep{AraqueParpuraEJN1998a,LiuNedergaard2004,Liu_PNAS2004}, or
release-increasing (\figref{fig:pre-loop}E), when astrocytic
glutamate binds NMDARs or group~I mGluRs
\citep{FiaccoMcCarthy2004,JourdainVolterra2007,Navarrete_Neuron2010,Bonansco_etal_EJN2011,Navarrete_PB2012,Perea_NatComm2014,Martin_Science2015}. The
functional implications of these modulations of synaptic release by
glutamatergic gliotransmission on synaptic transmission have been
widely addressed in a series of previous studies
\citep{DePitta_PCB11,DePitta_FCN13,DePitta_Neurosci2015}, and the
remainder of this section reviews and extends the main results from those
studies about short-term synaptic plastic and synaptic filtering.

\figref{fig:pre-loop}D (\textit{left panel}) shows how postsynaptic
currents (PSCs) change in the presence of release-decreasing
glutamatergic gliotransmission when elicited by two consecutive action
potentials arriving to the resting synapse 20~ms after the
onset of gliotransmission at $t=5$~s (\figref{fig:pre-loop}C). Two differences with
respect to the case without gliotransmission (\textit{black trace})
may be observed. First the PSC amplitude overall decreases
(\textit{red trace}), consistent with a decrease of synaptic efficacy
caused by the reduction of synaptic release by astrocytic
glutamate. Then, the second PSC is larger then the first one, which is
the opposite of what would be measured in the absence of
gliotransmission. In other words, in agreement with experimental
observations \citep{LiuNedergaard2004}, the release-decreasing effect
of astrocytic glutamate results in an increased pair pulse ratio (PPR)
with respect to the case without gliotransmission (PPR$_0$). Notably,
as shown in \figref{fig:pre-loop}D (\textit{right panel}), this change
in the PPR ratio is only transient and vanishes together with the
effect of gliotransmission on synaptic release. Similar considerations
also hold in the case of a release-increasing effect of astrocytic
glutamate on synaptic transmission \citep{JourdainVolterra2007}: while
PSC amplitude increases (\figref{fig:pre-loop}F, \textit{left panel},
\textit{green trace}), this occurs to the detriment of PPR, which
decreases instead (\figref{fig:pre-loop}F, \textit{right panel}).
Thus, synapses whose release probability is increased by glutamatergic
gliotransmission are likely to run out faster of neurotransmitter,
exhibiting rapid onset of short-term depression, consistent with lower
PPR values. On the contrary, synapses whose release probability is
reduced by astrocyte-released glutamate, deplete their
neurotransmitter resources slower and may exhibit progressive
facilitation (i.e. potentiation) of their efficacy to transmit
action potentials, and so larger PPR values
\citep{Dittman_etal_JN2000}. That is, the plasticity mode of a
synapse, namely whether it is depressing or facilitating, may not be
fixed but rather be modulated by glutamatergic gliotransmission by
surrounding astrocytes in an activity-dependent fashion
\citep{DePitta_PCB11,DePitta_FCN13}.

An important consequence of short-term synaptic dynamics is that
synapses can act as filters
\citep{MarkramWangTsodyks_PNAS1998,FortuneTiNS2001,AbbottRegehr_Nat2004}. Hence,
modulations of synaptic dynamics by glutamatergic gliotransmission are
also expected to affect the synapse's filtering characteristics
\citep{DePitta_Neurosci2015}. This scenario is illustrated in
\figref{fig:syn_filtering} where the effect of release-decreasing
vs. release-increasing glutamatergic gliotransmission, respectively on
depressing and facilitating synapses, is shown in terms of changes of
the filtering characteristics of these synapses, i.e. their
steady-state release as function of the frequency of presynaptic
stimulation \citep{AbbottRegehr_Nat2004}. In the absence of
gliotransmission, depressing synapses, which are characterized by
intermediate-to-high initial probability of release
\citep{Dittman_etal_JN2000} (\figref{fig:syn_filtering}A,
\textit{black circles}), predominantly act as low-pass filters
(\figref{fig:syn_filtering}B, \textit{black circles}) that are most
effective at transmitting low frequency pre-synaptic spike trains
(\figref{fig:syn_filtering}C, \textit{black traces}). On the contrary,
facilitating synapses, with a low-to-intermediate initial probability
of neurotransmitter release \citep{Dittman_etal_JN2000}
(\figref{fig:syn_filtering}A, \textit{black circles}), function as
high-pass or band-pass filters (\figref{fig:syn_filtering}B,
\textit{black circles}), that is they are mostly effective at
transmitting action potentials in an intermediate range of presynaptic
activity (\figref{fig:syn_filtering}C, \textit{black trace}).

In the presence of glutamate release by the astrocyte, these two
scenarios could be reversed. Consider indeed the simple heterosynaptic
case where glutamatergic gliotransmission is stimulated by other means than by the very synapses it impinges on. It may be noted that release-decreasing gliotransmission flattens the synaptic steady-state
release towards zero for all frequencies of stimulation
(\figref{fig:syn_filtering}B, \textit{red circles}), ensuing in synaptic
transmission that resembles the one of a facilitating, band-pass
synapse (compare the \textit{red}~PSC \textit{trace} in
\figref{fig:syn_filtering}C with the \textit{black}~PSC \textit{trace}
in \figref{fig:syn_filtering}F). Vice versa, release-increasing
gliotransmission could turn band-pass features of transmission by a
facilitating synapse (\figref{fig:syn_filtering}E, \textit{green
  circles}) into low-pass, reminiscent of a more depressing synapse
(compare the \textit{green}~PSC \textit{trace} in
\figref{fig:syn_filtering}F with the \textit{black}~PSC \textit{trace}
in \figref{fig:syn_filtering}C). On the other hand, when
gliotransmission is stimulated by the same synapses that it modulates -- that is, in the homosynaptic scenario of gliotransmission --, inspection
of the ensuing synaptic filtering characteristics
(Figure~\ref{fig:syn_filtering}B,E, \textit{cyan circles}) reveals
that these latter coincide with those obtained in the absence of
gliotransmission for low frequencies of presynaptic activity, while
they tend to equal those observed with heterosynaptic gliotransmission
as the frequency of stimulation increases. This coexistence of mixed
features from apparently opposite scenarios, i.e. no gliotransmission
vs. heterosynaptic gliotransmission, can be explained by the fact that
the release of glutamate from the astrocyte requires
intracellular~Ca$^{2+}$ to cross a threshold concentration. Hence, in
the homosynaptic scenario, synapses that impinge on the astrocyte must
be stimulated at rate sufficiently high to allow astrocytic~Ca$^{2+}$
to increase beyond such a threshold.

The modulation of synaptic filtering by glutamatergic gliotransmission
opens to the possiblity that the same stimulus could be differently
filtered (i.e. processed) and transmitted by a synapse in the presence
(or not) of glutamate release by surrounding astrocytic processes,
ultimately endowing that synapse with processing versatility with
respect to incoming action potentials. Moreover, to the extent that
synaptic dynamics critically shapes the computations performed by the
neural circuitry, such versatility could also be reflected at the
network level, leading to the possibility that the same neuron-glia
network could be involved in different computational tasks defined, time by time, by activity-dependent gliotransmitter release by astrocytes in the network.

\subsection*{Astrocyte-mediated slow inward currents}

Induction of slow inward (i.e. depolarizing) currents (SICs) by
activation of extrasynaptically-located postsynaptic NMDA receptors is
the other mechanism considered in this study whereby glutamatergic
gliotransmission could affect synaptic information transfer. While
astrocyte-mediated~SICs have been reported in several brain regions,
the pathway underlying glutamate release by astrocytes has not been
fully elucidated
\citep{AgulhonMcCarthy_NeuronRev2008,Papouin_PTRSB2014}. It is likely
that, similar to the presynaptic route for glutamatergic
gliotransmission discussed above, multiple pathways for glutamate
release could be used by the same astrocyte
\citep{ParpuraZorec_BRR2010}, but their deployment depends on
developmental, regional and physiological factors
\citep{Halassa_TMM2007}. Astrocytic~Ca$^{2+}$ activity seems a crucial
factor in the regulation of astrocyte-mediated~SICs
\citep{ParriCrunelli2001,Angulo2004,FellinCarmignoto_Neuron2004,PereaAraque_JN2005,DAscenzo_PNAS2007,Bardoni_JP2010,Pirttimaki_JN2011}. In
particular,~SIC frequency and amplitude have been shown to increase
upon~Ca$^{2+}$ elevations mediated by GPCRs on astrocytes such as
mGluRs
\citep{ParriCrunelli2001,Angulo2004,FellinCarmignoto_Neuron2004,PereaAraque_JN2005,DAscenzo_PNAS2007,Navarrete_CC2012,Navarrete_PB2012},
the metabotropic purinergic P2Y1 receptor \citep{Bardoni_JP2010}, the
endocannabinoid CB1 receptor \citep{NavarreteAraque_Neuron2008} or the
protease-activated receptor~1 (PAR1)
\citep{Shigetomi_JN2008}. Remarkably, stimulation of~PAR1s on
hippocampal astrocytes was shown to trigger, under physiological
conditions,~Ca$^{2+}$-dependent glutamate release from these cells
through Bestrophin-1 anion channel \citep{Oh_MB2012,Woo_Cell2012}, and
this pathway of glutamate release has been suggested as a candidate
mechanism for~SICs \citep{Papouin_Cell2012}. Channel-mediated
glutamate release is expected to account for prolonged ($>$10~s)
release of transmitter but in small amounts per unit time
\citep{Woo_Cell2012} thus ensuing in modest, very slow rising and
decaying inward currents. While similar~SICs have indeed been recorded
\citep{AraqueParpuraEJN1998a,Lee_JP2007}, most experiments
reported~SICs within a wide range of amplitudes to last only few
seconds at most and, rise in correlation with astrocytic~Ca$^{2+}$
increases with rise time much shorter than their decay
\citep{FellinCarmignoto_Neuron2004,Angulo2004,PereaAraque_JN2005,Shigetomi_JN2008,Nie_JNP2010,ReyesHaro_JGP2010,Chen_PNAS2012,Martin_Science2015}
akin to currents that would ensue from a quantal mechanism of
gliotransmitter release \citep{Sahlender_PTRSB2014}.

Based on these arguments, we assume glutamate exocytosis as the
candidate mechanism for glutamate release by astrocytes that mediates~SICs. Accordingly, we adopt the description of astrocytic glutamate exocytosis previously introduced 
(Figures~\ref{fig:tripartite_io}G--I) to also model
astrocyte-mediated~SICs. In this fashion, glutamate exocytosis by
the astrocyte into the extracellular space (\figref{fig:sics}A) results 
in activation of extrasynaptically-located NMDARs on nearby neuronal dendrites which trigger~SICs 
(\figref{fig:sics}B) that cause slow depolarizing postsynaptic potentials (PSP, \figref{fig:sics}C).

An important functional consequence of~SIC-mediated depolarizations,
is that they can modulate neuronal excitability
\citep{FellinCarmignoto_Neuron2004,PereaAraque_JN2005,DAscenzo_PNAS2007,Nie_JNP2010}. As
illustrated in Figures~\ref{fig:sics}D,E, astrocyte-mediated~SICs
(\textit{cyan trace}) may add to regular synaptic currents
(\textit{black trace}) resulting in depolarizations of postsynaptic
neurons closer to their firing threshold \citep{DAscenzo_PNAS2007}. In
turn, these larger depolarizations could dramatically change
generation and timing of action potentials by those neurons in
response to incoming synaptic stimuli (\figref{fig:sics}F). This could
ultimately affect several neurons within the reach of glutamate released
by an astrocyte, leading to synchronous transient increases of their
firing activity \citep{FellinCarmignoto_Neuron2004}. Remarkably, this
concerted increase of neuronal excitability has often been observed in
correspondence with large amplitude (i.e. $>$100~pA)~SICs
\citep{FellinCarmignoto_Neuron2004,KangKang2005,Bardoni_JP2010,Nie_JNP2010},
but experiments report the majority of~SICs to be generally smaller,
with amplitudes $<$80~pA
\citep{FellinCarmignoto_Neuron2004,KangKang2005,PereaAraque_JN2005,Chen_PNAS2012,Perea_NatComm2014,Martin_Science2015}.
It is therefore unclear whether SIC-mediate increase of
neuronal excitability could occur \citep{FellinCarmignotoHaydon2006}
or not \citep{KangKang2005,TianNedergaard_Nature2005,Ding_JN2007} in
physiological conditions.

In Figure~\ref{fig:sics}G, we consider postsynaptic firing in a
standard leaky integrate and fire neuron model
\citep{Fourcaud_NC2002,Burkitt_BC2006} as a function of presynaptic
activity for~SICs of different amplitudes (30--45~pA, see
\appref{app:par-estimation}) randomly occurring at an average rate of
1~Hz based on a binomial process for glutamate release from astrocytes
as suggested by experiments \citep{Santello_etal_Neuron2011} (see
\appref{app:model}). In line with experimental evidence
\citep{Rauch_JNP2003}, the input-output transfer function in the absence of gliotransmission has a typical sigmoidal shape (\textit{black dots}) which reflects: (i)~gradual emergence of firing for
low ($>$10~Hz) fluctuating synaptic inputs; (ii)~the progressive,
quasi-linear increase of the firing rate for presynaptic activity
beyond $\sim$30~Hz; and finally, (iii)~saturation of the firing rate for
sufficiently strong synaptic inputs such that timing of action
potential generation approaches the neuron's refractory period (which
was fixed at 2~ms in the simulations, \appref{app:par-estimation})
\citep{Burkitt_BC2006}. The addition of astrocyte-mediated~SICs alters
the firing characteristics of the neuron due to the ensuing larger
depolarization. In particular the neuron could generate action
potentials for lower rates of presynaptic activity
(\textit{cyan}/\textit{blue dots}). Clearly, the larger the~SIC is, the more postsynaptic firing increases with respect to the case without SICs, for a given level of presynaptic activity.

As previously mentioned, these results assume an average 1~Hz rate for
astrocyte-mediated~SICs. While this is possible in principle, it seems
unlikely as following explained. The weak correlation of~SIC amplitude
with somatic~Ca$^{2+}$ elevations observed in experiments favors
indeed the idea that glutamate-mediate~SICs are highly localized
events, occurring within subcellular domains at astrocytic processes
\citep{PereaAraque_JN2005}. In turn,~Ca$^{2+}$-elevations in
astrocytic processes could be as short-lived as $\sim$0.5~s
\citep{DiCastro_Volterra_NatNeurosci2011,Panatier_etal_Cell2011}, thus
in principle allowing for glutamate release rates of the order
of 1 Hz. However, in practice, reported~SIC frequency are much lower,
that is $<$5/min (i.e.~$\sim$0.08~Hz)
\citep{PereaAraque_JN2005,Perea_NatComm2014}. Hence, it may be
expected that the effect of~SICs on neuronal firing could be
considerably reduced with respect to the case considered in
\figref{fig:sics}G. 

We consider this possibility more closely in \figref{fig:sics}H, where we analyze postsynaptic firing in function of the average frequency of astrocyte-mediated~SICs, both in the absence of synaptic activity
(\textit{black} and \textit{dark blue dots}), and in the case of
presynaptic activity at an average rate $\sim$1~Hz, which corresponds to
typical levels of spontaneous activity in vivo \citep{Hromadka_PB2008} (\textit{grey} and \textit{light blue
  dots}). It may be noted that the effect of~SICs of typical
amplitudes on postsynaptic firing rate is generally small, i.e. $<$0.5~Hz,
except for unrealistic ($>$0.1~Hz)~SIC rates, while it gets
stronger in association with synaptic activity. In this latter case
however, the possible increase in postsynaptic firing by
astrocyte-mediated~SICs, is limited by the rate of reintegration of
released glutamate resources in the astrocyte (fixed at $\sim$1~Hz,
\appref{app:par-estimation}). Analogously to short-term synaptic
depression in fact, our description of gliotransmitter release
predicts that for release rates that exceed the rate of reintegration
of released glutamate by the astrocyte, exhaustion of astrocytic
glutamate resources available for further release will result in~SICs
of smaller amplitude. In this fashion, due to depletion of astrocytic
glutamate, the effect of large rates of glutamate release, and
thus of~SICs, on neuronal firing tends to be equivalent to that of
considerably lower ones. 

Taken together, the above results do not exclude a possible role
of~SICs in modulation of neuronal excitability and firing but suggest
that such modulation could effectively occur only in coincidence with
proper levels of synaptic activity. In this fashion,
astrocyte-mediated~SICs could be regarded to operate a sort of
coincidence detection between synaptic activity and astrocytic
glutamate release \citep{PereaAraque_JN2005}, whose readout would then
be a temporally precise, cell-specific increase of neuronal firing
(\figref{fig:sics}F).

\section*{Astrocyte-mediated regulation of long-term plasticity}

The strength of a synaptic connection between two neurons can be
modified by activity, in a way that depends on the timing of neuronal
firing on both sides of the synapse, through a series of processes
collectively known as spike-timing--dependent plasticity (STDP)
\citep{Caporale_ARN2008}. As both pre- and postsynaptic pathways of
glutamatergic gliotransmission potentially change EPSC magnitude, thereby affecting postsynaptic firing, it may be expected that they could also influence~STDP.

Although the molecular mechanisms of STDP remain debated, and
different mechanisms could be possible depending on type of synapse,
age, and induction protocol \citep{Froemke_FSN2010}, at several
central excitatory synapses postsynaptic calcium concentration has
been pointed out as a necessary factor in induction of synaptic
changes by STDP
\citep{Magee_Science1997,Ismailov_JN2004,NevianSakmann_JN2004,Bender_JN2006,Nevian_JN2006}. Remarkably,
amplitude and, likely, time course of postsynaptic~Ca$^{2+}$ could
control the direction of plasticity: smaller, slower increases of
postsynaptic~Ca$^{2+}$ give rise to spike-timing--dependent long-term
depression (LTD) whereas larger, more rapid increases cause
spike-timing--dependent long-term potentiation (LTP)
\citep{Magee_Science1997,Ismailov_JN2004,Nevian_JN2006}. In
calcium-based~STDP models, this is also known as the
``Ca$^{2+}$-control hypothesis''
\citep{Shouval_PNAS2002,Cai_JNP2007,Graupner_FCN2010}. According to
this hypothesis, no modification of synaptic strength occurs when~Ca$^{2+}$ is below a threshold $\theta_d$ that is larger than the resting~Ca$^{2+}$ concentration. If calcium resides in an
intermediate concentration range, between $\theta_d$ and a second
threshold $\theta_p > \theta_d$, the synaptic strength is
decreased. Finally, if calcium increases above the second threshold,
$\theta_p$, the synaptic strength is potentiated.

Figures~\ref{fig:stdp-curves}A.1 and~\ref{fig:stdp-curves}B.1 exemplify the operational
mechanism of the~Ca$^{2+}$-control hypothesis within the framework of
a nonlinear~Ca$^{2+}$-based model for~STDP at glutamatergic synapses
originally introduced by \citet{Graupner_PNAS2012}. At most
glutamatergic synapses, postsynaptic~Ca$^{2+}$ is mainly regulated by
two processes: (i)~postsynaptic~Ca$^{2+}$ entry mediated by NMDARs
\citep{Malenka_Neuron2004}, and~(ii)~Ca$^{2+}$ influx by
voltage-dependent~Ca$^{2+}$ channels (VDCCs)
\citep{Magee_CON2005,Bender_JN2006,Nevian_JN2006,Sjostrom_PR2008}. In
this fashion, each presynaptic action potential generates a
long-lasting~Ca$^{2+}$ transient by opening NMDAR channels, while
postsynaptic firing results in a short-lasting~Ca$^{2+}$ transient due
to opening of VDCCs by dendritic depolarization through
back-propagating action potentials~(bAPs)
\citep{Caporale_ARN2008}. Presynaptic action potentials alone do not
trigger changes in synaptic strength, but they do so in correlation
with postsynapitc bAPs \citep{SjostromNelson_CON2002}. Notably \citep{AbbottNelson2000}, in a
typical~STDP induction pairing protocol,~LTD
is induced if the postsynaptic neuron fires before the presynaptic
one, i.e. post$\rightarrow$pre pairing at negative spike timing
intervals $\Delta t$ (Figures~\ref{fig:stdp-curves}A.1). Contrarily, LTP~is induced when the presynaptic cell fires before the postsynaptic cell, that is for pre$\rightarrow$post pairing at positive $\Delta t$
intervals (Figures~\ref{fig:stdp-curves}A.1). This is possible
because, when a presynaptic action potential is followed shortly after
by a postsynaptic~bAP, the strong depolarization by this latter drastically increases the voltage-dependent NMDAR-mediated~Ca$^{2+}$ current due to removal of the NMDAR magnesium block
\citep{Nowak1984,JahrStevens1990}, thereby resulting in supralinear
superposition of the NMDAR- and VDCC-mediated~Ca$^{2+}$
influxes. 

In the framework of the~Ca$^{2+}$-control hypothesis, these
observations may be summarized as follows. For large $\Delta t$, pre-
and postsynaptic Ca$^{2+}$ transients do not interact, and the
contributions from potentiation and depression by pre/post pairs (or
vice versa) cancel each other, leading to no synaptic changes on
average (\figref{fig:stdp-curves}C, \textit{black curves}). For short, negative $\Delta t$, the presynaptically evoked Ca$^{2+}$ transient rises instead above the depression threshold ($\theta_d$) but not
beyond the potentiation threshold ($\theta_p$). Consequently,
depression increases whereas potentiation remains constant, which
leads to~LTD induction. For short, positive $\Delta t$ however, the
postsynaptically evoked calcium transient rises on top of the
presynaptic transient by the NMDAR nonlinearity, and increases
activation of both depression and potentiation. Because the rate of
potentiation is larger than the rate of depression
(\appref{TA:All-Parameters}), this results in~LTP induction.

For the same number of pre/post pairs (or vice versa), mapping of the
average synaptic modification as function of the spike timing interval $\Delta t$, ultimately provides an~STDP curve that qualitatively resembles the classic curve originally described by
\citet{BiPoo_JN1998} (\figref{fig:stdp-curves}C, \textit{bottom
  panel}, \textit{black curve}). In the following, we will focus on
parameters that lead to such a STDP curve and investigate how this
curve is affected in the presence of glutamatergic gliotransmission,
through the pre- and postsynaptic pathways of regulation discussed
above.

\subsection*{Presynaptic pathway}

The very nature of synaptic transmission crucially depends on the
synapse's initial probability of neurotransmitter release insofar as
this latter sets both the tone of synaptic transmission, that is how
much neurotransmitter is released per action potential by
the synapse on average, as well as whether the synapse displays short-term
depression or facilitation \citep{AbbottRegehr_Nat2004}. Synapses with
low-to-intermediate values of initial neurotransmitter release
probability, like for example, Schaffer collateral synapses
\citep{Dittman_etal_JN2000}, or some cortical synapses
\citep{MarkramWangTsodyks_PNAS1998}, are indeed prone to display
facilitation, whereas synapses that are characterized by large release
probability are generally depressing
\citep{MarkramWangTsodyks_PNAS1998}. Because synaptic release
probability also dictates the degree of activation of~NMDARs, and
consequently, the magnitude of postsynaptic~Ca$^{2+}$ influx, it is
expected that both the tone of synaptic transmission and its short-term
dynamics could affect~STDP \citep{Froemke_FSN2010}. The relative
weight of these two factors in shaping synaptic changes however,
likely depends on the protocol for~STDP induction. Short-term
plasticity could indeed sensibly regulate~STDP induction only for
rates of presynaptic action potentials high enough to allow
facilitation or depression of synaptic release from one~AP to the
following one \citep{Froemke_Nature2002,Froemke_JNP2006}. In this
study, we consider low pre/post frequencies of~1~Hz. At such
frequencies we expect short-term plasticity to have a negligible
effect, and thus we only focus on how changes in the tone of synaptic
transmission by glutamatergic gliotransmission affect STDP.

Figures~\ref{fig:stdp-curves}A.2,B.2 respectively show the
outcome of~LTD and~LTP induction for two consecutive
pre$\rightarrow$post and pre$\rightarrow$post pairings preceded  by the onset of release-decreasing gliotransmission at 0.1~s (\textit{top panels}, \textit{black marks}). A comparison of the ensuing
postsynaptic~Ca$^{2+}$ dynamics with respect to the case without
gliotransmission (Figures~\ref{fig:stdp-curves}A.1,B.1) reveals that
the strong decrease of synaptic release probability (S.R.P.,
\textit{top panels}, \textit{red curves}) caused by gliotransmission
remarkably reduces the NMDAR-mediated contribution to
postsynaptic~Ca$^{2+}$ influx (\textit{middle panels}), resulting in
smaller variations of synaptic strength (\textit{bottom panels}). In
this fashion, at the end of the pairing protocol, release-decreasing
gliotransmission accounts for less time spent
by~Ca$^{2+}$ above either thresholds of~LTD and~LTP
(\figref{fig:stdp-curves}C, \textit{top panel}, \textit{red
  traces}). The resulting STDP curve thus displays strongly attenuated
LTD and LTP (\figref{fig:stdp-curves}C, \textit{bottom panel},
\textit{red curve}), with windows for these latter spanning a
considerably smaller range of $\Delta t$s than in the curve obtained
without gliotransmission (\textit{black curve}).

Similar considerations apply to the case of release-increasing
gliotransmission (Figures~\ref{fig:stdp-curves}A.3,B.3). In this case,
the NMDAR component of postsynaptic~Ca$^{2+}$ could increase by
gliotransmission even beyond the $\theta_d$ threshold (\textit{dashed
  blue line}), thus favoring depression while reducing potentiation
(\textit{bottom panels}). In particular, for short positive $\Delta
t$, the maximal~LTP does not change but the $\Delta t$ range for~LTP
induction shrinks. For $\Delta t>40$~ms in fact, the time
that~Ca$^{2+}$ spends above the~LTD threshold increases with respect to
the time spent by~Ca$^{2+}$ above the~LTP threshold,
thereby resulting in~LTD induction (Figures~\ref{fig:stdp-curves}C,
\textit{top panel}, \textit{green traces}). In this fashion, the~STDP
curve in the presence of release-increasing gliotranmission displays a
narrow 0--40~ms~LTP window outside which~LTD occurs instead
(Figures~\ref{fig:stdp-curves}C, \textit{bottom panel}, \textit{green
  curve}).

\figref{fig:stdp-curves}D summarizes how the~STDP curve changes for
the whole spectrum of glutamatergic gliotransmission. In this figure,
a y-axis value of ``Gliotransmission Type'' equal to~0 corresponds to
maximum release-decreasing gliotransmission (\textit{red curve} in \figref{fig:pre-loop}C); a value equal to~1
stands instead for maximum release-increasing gliotransmission (as in
the case of the \textit{green curve} in \figref{fig:pre-loop}C); finally, a value of~0.5 corresponds to no effect of gliotransmission on synaptic release (\textit{black curve} in \figref{fig:pre-loop}C). It
may be noted that gliotransmission may affect the~STDP curve in
several ways, changing both strength of plastic changes (\textit{color
  code}) as well as shape and areas of~LTP and~LTD windows. In
particular, as revealed by \figref{fig:stdp-curves}E, maxima of~LTP
(\textit{cyan circles}) and~LTD (\textit{yellow circles}) decrease
with decreasing values of gliotransmission type, consistently with
smaller postsynaptic~Ca$^{2+}$ influx for larger decreases of synaptic
release by gliotransmission. This suggests that release-decreasing
gliotransmission (\textit{red-shaded area}) could attenuate~STDP yet
in a peculiar fashion, counteracting~LTD more than~LTP induction, as
reflected by increasing values of~LTP/LTD area ratio (\textit{magenta
  curve}).

On the contrary, the effect of release-increasing gliotransmission
(\figref{fig:stdp-curves}E, \textit{green-shaded area}) could be
dramatically different. For sufficiently strong
increases of synaptic release by gliotransmission in fact, the LTP/LTD area
ratio drops to zero (\textit{hatched area}) in correspondence with the
appearance of two ``open'' LTD windows, one for $\Delta t<0$ and the
other for sufficiently large positive spike timing intervals. In
parallel, consistently with the fact that release-increasing
gliotransmission tends to increase the fraction of time spent by
postsynaptic Ca$^{2+}$ above the threshold for LTD thereby promoting
this latter (\figref{fig:stdp-curves}C), the range for LTP induction
also tends to shrink to lower $\Delta t$ values as release-increasing
gliotransmission grows stronger (\figref{fig:stdp-curves}D,
\textit{red color-coded areas} for Gliotransmission Type $>$0.5).

In summary, our analysis reveals that modulation of synaptic release
by glutamatergic gliotransmission could change STDP both
quantitatively and qualitatively, from hindering its induction for
release-decreasing modulations, to altering both shape and existence
of~LTD windows for release-increasing modulations. However, whether and how
this could effectively be observed in experiments remains to
be investigated. Supported both by experimental evidence and
theoretical arguments is the notion that regulations of the tone of
synaptic transmission by glutamatergic gliotransmission likely require
specific morphological and functional constraints to be fulfilled by the nature of astrocyte-synapse coupling
\citep{Araque_Neuron2014,DePitta_Neurosci2015}. Similar arguments
may ultimately hold true also for modulation of STDP, insofar as for
this modulation to be measured in our simulations, we required both a
sufficiently strong increase/decrease of synaptic release by
gliotransmission and a decay time of such increase/decrease long
enough for this latter to be present during the
induction protocol. Should these two aspects not have been fulfilled
in our simulations, then modulations of STDP by
gliotransmitter-mediated changes of synaptic release would likely have
been negligible or even undetectable.

\subsection*{Postsynaptic pathway}

We now turn our analysis to the possible impact of
astrocyte-mediated~SICs on~STDP. Because~SICs are through
extrasynaptic~NMDA receptors and these receptors are mainly permeable
to~Ca$^{2+}$ ions \citep{Cull-Candy_CON2001}, then~SICs could
contribute to postsynaptic~Ca$^{2+}$ thereby affecting~STDP. Nevertheless, we should note that
it is unclear whether and how extrasynaptic~NMDARs contribute to
plasticity, independently of the occurrence of SICs
\citep{Papouin_PTRSB2014}. For example, theta-burst LTP induction in~CA1 neurons of
rat hippocampal slices, is turned into LTD when extracellular NMDARs
are selectively stimulated 
\citep{Liu_BBR2013}, but it is unknown whether these receptors
have a role in STDP \citep{Evans_BB2015}. In general, for a given STDP
induction protocol, two factors that could crucially regulate
how~Ca$^{2+}$ transients mediated by extrasynaptic NMDARs are involved
in STDP, are the location of these receptors on the spine and the
morphology of this latter in terms of spine head and neck
\citep{Bloodgood_CON2007,Rusakov_Science2004}. Unfortunately both
these factors remain unknown in the current knowledge of SIC-mediating
extrasynaptic NMDARs and, for the remainder of this study, we assume
that, in spite of their possible location away from the postsynaptic
density along the spine neck or the dendritic shaft
\citep{Petralia_Neuroscience2010}, SIC-mediating extrasynaptic NMDARs
could still regulate spine~Ca$^{2+}$ dynamics \citep{Halassa_TMM2007}.

Based on the above rationale, we thus model~SICs as slow
potsynaptic~Ca$^{2+}$ transients that will add to presynaptically- and
postsynaptically-triggered ones, and study their effect on the
induction of SDTP by classic pairing protocols. For the sake of
generality, we express the peak of~SIC-mediated~Ca$^{2+}$ transients
in units of NMDAR-mediated EPSCs. However, since in our~STDP
description individual EPSCs do not trigger any synaptic modification
\citep{Graupner_PNAS2012}, then we may expect that only~SICs sufficiently
larger than EPSCs could effectively affect~STDP. On the other hand,
smaller~SICs could also sum with~Ca$^{2+}$ transients by pre/post
pairings resulting in~Ca$^{2+}$ elevations beyond either~LTD or~LTP
thresholds that would ultimately cause synaptic changes
(Figures~\ref{fig:stdp_sic}A,B). Hence, based on these considerations, we deem amplitude and timing of~SICs, both in terms of frequency of occurrence and onset with respect to~STDP-inducing
stimuli, to be crucial factors in shaping how~SICs affect~STDP, and
thus we set to analyze these three factors separately.

Figure~\ref{fig:stdp_sic}C summarizes the
results of our simulations for~SICs as large as 0.5,~1 or 1.5~times
typical EPSCs, occurring at a fixed rate of 0.1~Hz and starting 100~ms
before the delivery of 60~STDP-inducing pre/post pairings at 1~Hz. As illustrated in
Figures~\ref{fig:stdp_sic}A,B, for the same~SIC kinetics, these
simulations guarantee superposition between~Ca$^{2+}$ influxes
mediated by~SICs and pre/post pairings such that the extension of the
ensuing~Ca$^{2+}$ transient beyond~LTD and~LTP thresholds
(\textit{dashed lines}) merely depends on~SIC amplitude. In this
fashion, it may be noted that~SICs of amplitude smaller than or equal to
typical EPSCs (\figref{fig:stdp_sic}C, \textit{turquoise circles} and \textit{black circles} respectively), that alone would not produce any synaptic modification, do not sensibly change the~STDP curve with
respect to the previously considered case of an alike synapse in the
absence of gliotransmission (\figref{fig:stdp-curves}C, \textit{black
  circles}). Conversely, large~SICs could dramatically affect~STDP,
shifting the~STDP curve towards negative synaptic changes
(\textit{blue circles}), and this negative shift increases the
larger~SICs grow beyond the $\theta_d$ threshold (results not shown). In
this case,~STDP generally results in~LTD with the exception of a~LTP
window that is comprised between $\sim$0~ms and positive $\Delta t$
values that are smaller than those in the absence of gliotransmission
(\figref{fig:stdp-curves}C, \textit{green circles}). This resembles
what previously observed for~STDP curves in the presence of
release-increasing gliotransmission, with the only difference that,
for large $|\Delta t|$ values,~LTD strength in the presence of
astrocyte-mediated~SICs is found to be the same, regardless of $\Delta
t$ (compare the \textit{blue curve} in \figref{fig:stdp_sic}C with the
\textit{green curve} in \figref{fig:stdp-curves}C).

In Figure~\ref{fig:stdp_sic}D we consider the alternative scenario where
only~SICs as large as typical EPSCs impinge on the postsynaptic neuron
at different rates, yet always 100~ms before~STDP-inducing
pairings. Akin to what happens for~SIC amplitudes, the larger the~SIC frequency is, the more the~STDP curve changes. Indeed, as~SIC frequency increases above~SIC decay rate (i.e. $1/\tau_A$, \appref{app:model}, \secref{app-sec:glt-rel}),~SIC-mediated~Ca$^{2+}$ transients start adding up, so that the fraction of time spent
by~Ca$^{2+}$ beyond the~LTD threshold increases favoring~LTD
induction. In this fashion, the ensuing~STDP curve, once again,
consists of a narrow~LTP window for $\Delta t\ge 0$, outside which
only~LTD is observed (\textit{red curve}). In practice however,
because~SICs occur at rates that are much slower than their typical
decay (\appref{app:par-estimation}), they likely affect~STDP in a more
subtle fashion. This may be readily understood considering the
\textit{pink}~STDP \textit{curve} obtained for~SICs at 0.1~Hz, that is
the maximum rate experimentally recorded for these currents
\citep{PereaAraque_JN2005}. Inspection of this curve indeed suggests
that~SICs could effectively modulate~LTD and~LTP maxima as well as the
outer sides of the~LTD/LTP windows, which dictate how fast
depression/potentiation decay for large $|\Delta t|$, but overall the
qualitative features of the~STDP curve are preserved with respect to
the case without gliotransmission (\textit{black curve}).

Clearly, the extent of the impact of~SIC amplitude and frequency
on~STDP discussed in Figures~\ref{fig:stdp_sic}C,D ultimately depends
on when~SICs occur with respect to ongoing~STDP-inducing pairings. Had
we set~SICs to occur $\sim$200~ms after pre/post~Ca$^{2+}$ transients
in our simulations, then, as illustrated in
Figures~\ref{fig:stdp_sic}E,F, we would have not detected any sensible
alteration of~STDP, unless~SICs were larger than typical EPSCs and/or
occurred at sufficiently high rate to generate~Ca$^{2+}$ transients
beyond the plasticity thresholds (results not shown). To seek
understanding of how timing of~SICs vs.~pre/post pairings could alter~LTD and/or~LTP,
we simulated~STDP induction by pairing as the time interval ($\Delta
\varsigma$) between~SIC and pre/post pairs was systematically varied
(with~SIC rate fixed at 0.2~Hz) (Figures~\ref{fig:stdp_sic}G--I). In
doing so, we adopted the convention that negative $\Delta \varsigma$
values stand for~SICs preceding pre/post (or post/pre) pairings while,
positive $\Delta \varsigma$ refer to the opposite scenario of~SICs
that follow pairings (\figref{fig:stdp_sic}G, \textit{top
  schematics}). Then, it may be observed that, for $\Delta \varsigma$
in between approximately -300~ms and 0~ms,~LTD could be induced for
any negative $\Delta t$ as well as for large positive $\Delta t$
(\figref{fig:stdp_sic}G, \textit{blue tones}) -- in this latter case
to the detriment of the~LTP window, whose upper bound moves to lower
$\Delta t$ values (\figref{fig:stdp_sic}G, \textit{red tones}). This
results in~STDP curves (e.g. \figref{fig:stdp_sic}J, \textit{yellow
  curve} for $\Delta \varsigma=-75$~ms) that bear strong analogy with
the blue and red curves in Figures~\ref{fig:stdp_sic}C,D respectively
obtained for~SICs of large amplitude and frequency, and suggest that depression grows as~SICs tend to concur with pre/post pairings. An inspection of postsynaptic~Ca$^{2+}$ transients
(Figures~\ref{fig:stdp_sic}H,I) indeed reveals that coincidence
of~SICs and pre/post pairings, which occurs at negative $\Delta
\varsigma$ of the order of~SIC rise time (see
\appref{app:par-estimation}), corresponds to the longest time spent
by~Ca$^{2+}$ above the~LTD threshold, thereby resulting in maximum~LTD (\figref{fig:stdp_sic}K) and thus, minimum~LTP
(\figref{fig:stdp_sic}L). Clearly, the $\Delta \varsigma$ range for
which coincidence of~SICs with pre/post pairings enhances~LTD
induction ultimately depends on kinetics of~SICs, as reflected by
their rise ($\tau_s^r$) and/or decay time constants ($\tau_s$), and
spans $\Delta \varsigma$ values approximately comprised between
$\pm$~SIC duration (i.e. $\simeq \tau_s^r + \tau_s$). As~SIC duration
increases in fact, either because of larger $\tau_s^r$ or larger
$\tau_s$ or both, so does the $\Delta \varsigma$ range for~LTD
enhancement, as reflected by the \textit{orange} and \textit{blue
  curves} in Figures~\ref{fig:stdp_sic}J--L.

In conclusion the simulations in Figures~\ref{fig:stdp_sic}G--L point
to both timing and duration of~SICs with respect to pre/post
pairing-mediated~Ca$^{2+}$ transients as a further,
potentially-crucial factor in setting strength and polarity of~STDP at
glutamatergic synapses. It is noteworthy to emphasize that, however, to
appreciate some effect on~STDP, we had to assume in those
simulations~SICs occurring at~0.2~Hz, that is two-fold the maximum~SIC
rate (i.e. $\sim0.1$~Hz) experimentally observed
\citep{PereaAraque_JN2005}. Indeed, analogous simulations run with
realistic~SIC rates $\le$0.1~Hz did produce only marginal changes
to~STDP curves, akin to those previously observed for the pink~STDP
curve in \figref{fig:stdp_sic}G. The potential functional implications
(or lack thereof) of this perhaps puzzling result are addressed in the following Discussion.
 
\section*{Discussion}

A large body of evidence has accumulated over the last years suggesting an active role of astrocytes in many brain
functions. Collectively, these data fuelled the concept that synapses
could be tripartite rather than bipartite, since in addition to the
pre- and postsynaptic terminals, the astrocyte could be an active
element in synaptic transmission
\citep{AraqueHaydonTripartite1999,Haydon2001,VolterraMeldolesiRev2005}. Using
a computational modeling approach, we showed here that glutamatergic
gliotransmission could indeed play several roles in synaptic
information transfer, either modulating synaptic filtering or
controlling postsynaptic neuronal firing, as well as regulating both
short- and long-term forms of synaptic plasticity. Supported by
experimental observations
\citep{LiuNedergaard2004,JourdainVolterra2007,DAscenzo_PNAS2007,Bonansco_etal_EJN2011,Perez_JN2014},
these results complement and extend previous theoretical work on
astrocyte-mediated regulations of synaptic transmission and plasticity
\citep{DePitta_FCN13,DePitta_Neurosci2015}, and pinpoint biophysical
conditions for a possible role of glutamatergic gliotransmission in
spike-timing--dependent plasticity.

An important prediction of our model indeed is that both pathways of
regulation of synaptic transmission by astrocytic glutamate considered
in this study -- presynaptic modulation of transmitter release and
postsynaptic~SICs -- could affect~STDP, potentially altering induction
of~LTP and~LTD. This alteration could encompass changes in the timing
between pre- and postsynaptic firing that is required for plasticity
induction, as well as different variations of synaptic strength in
response to the same stimulus. With this regard, the increase of LTP observed in our simulations, when moving from release-decreasing to release-increasing gliotransmission (\figref{fig:stdp-curves}E),
agrees with the experimental observation that~LTP induction at
hippocampal synapses requires weaker stimuli in the presence of
endogenous glutamatergic gliotransmission rather than when
gliotransmission is inhibited thereby decreasing synaptic release
probability \citep{Bonansco_etal_EJN2011}.

Notably, spike-timing--dependent plasticity in the hippocampus is not fully understood insofar as STDP induction by pairing protocols has produced a variety of seemingly contradicting observations for this brain region \citep{Buchanan_FSN2010}. Recordings in hippocampal slices for example, showed that pairing of single pre- and postsynaptic action potentials at positive spike timing intervals could trigger LTP \citep{Meredith_JN2003,Buchanan_JP2007,Campanac_JP2008}, as effectively expected by the classic STDP curve \citep{BiPoo_JN1998}, but also induce either LTD \citep{Wittenberg_JN2006} or no plasticity at all \citep{Buchanan_JP2007}. Although different experimental and physiological factors could account for these diverse observations \citep{Buchanan_FSN2010,ShulzJacob_FSN2010}, we may speculate that glutamatergic gliotransmission by astrocytes, which in those experiments was not explicitly taken into account, could also provide an alternative explanation. For example, the prediction of our model that release-increasing glutamatergic gliotransmission could account for multiple LTD windows, either at positive or negative spike timing intervals (\figref{fig:stdp-curves}), indeed supports the possibility that LTD in the hippocampus could also be induced by proper presentations of pre$\rightarrow$post pairings sequences \citep{Wittenberg_JN2006}. On the same line of reasoning, the possibility that astrocyte-mediated SICs could transiently increase postsynaptic firing (\figref{fig:sics}F), could explain why, in some experiments, precise spike timing in the induction of synaptic plasticity in the hippocampus could exist only
when single EPSPs are paired with postsynaptic bursts
\citep{Pike_JP1999,Wittenberg_JN2006}. Moreover, it was
also shown that postsynaptic firing is relatively less important than
EPSP amplitude for the induction of~STDP in the immature hippocampus
compared to the mature network, possibly due to a reduced
backpropagation of somatic APs in juvenile animals
\citep{Buchanan_JP2007}. Remarkably, these diverse modes of plasticity
induction could also ensue from different dynamics of glutamatergic
gliotransmission, as likely mirrored by the developmental profile of
somatic~Ca$^{2+}$ signals in hippocampal astrocytes
\citep{Volterra_NRN2014}, which have been reported to be much more
frequent in young mice \citep{Sun_Science2013}. Insofar as
somatic~Ca$^{2+}$ signals may result in robust astrocytic glutamate
release that could trigger, in turn, similar increases of synaptic
release and/or~SICs \citep{Araque_Neuron2014,Sahlender_PTRSB2014}, the
frequent occurrence of these latter could then ultimately guarantee a
level of dendritic depolarization sufficient to produce~LTP in mice
pups \citep{Golding_etal_Nature2002}.

High amplitude/rate~SICs, or large increases of synaptic release
mediated by glutamatergic gliotransmission, result, in our simulations,
in~LTD induction for any spike timing interval except for a
narrow~LTP window at small-to-intermediate $\Delta t>0$. This is in
stark contrast with~STDP experiments, where the observed plasticity
always depends, to some extent, on the coincidence of pre- and
postsynaptic activity, as EPSPs or postsynaptic action potentials fail
to induce plasticity by their own
\citep{SjostromNelson_CON2002,Caporale_ARN2008}. Apart from the
consideration that large~SIC amplitudes/rates and large increases of
synaptic release by astrocytic glutamate may not reflect physiological
conditions \citep{Ding_JN2007,AgulhonMcCarthy_NeuronRev2008}, this
contrast may be further resolved on the basis of the following
arguments.

A first consideration is that we simulated
plasticity induction assuming either persistent occurrence of~SICs or
continuous modulations of synaptic release during the whole induction
protocol. While this rationale proved useful to identify the possible
mechanisms of regulation of~STDP by glutamatergic gliotransmission, it
may likely not reflect what occurs in reality. Indeed, modulations of
synaptic release by glutamatergic gliotransmission could last only few
tens of seconds \citep{FiaccoMcCarthy2004,JourdainVolterra2007} and
thus be short-lived with respect to typical induction protocols which
are of the order of minutes
\citep{BiPoo2001,SjostromNelson_CON2002,Sjostrom_PR2008}. Moreover,
the morphology of astrocytic perisynaptic processes is not fixed but
likely undergoes dynamical reshaping in an activity-dependent fashion
during plasticity induction \citep{Lavialle_PNAS2011,Perez_JN2014},
thereby potentially setting time and spatial range of action of
gliotransmission on nearby synaptic terminals
\citep{DePitta_Neurosci2015}. In this fashion,~LTD for large spike
timing intervals could be induced only transiently and at selected
synapses, focally targeted by glutamatergic gliotransmission, while
leaving unchanged the qualitative features of the classic~STDP curve
obtained by somatic recordings in the postsynaptic neuron
\citep{BiPoo2001}.

A further aspect that we did not take
into account in our simulations is also the possible voltage
dependence of astrocyte-triggered~SICs. The exact nature of this
dependence remains to be elucidated and likely changes with subunit
composition of NMDA receptors that mediate~SICs in different brain
regions and at different developmental stages
\citep{Papouin_PTRSB2014}. Regardless, it may be generally assumed
that slow inward currents through NMDA receptors become substantial
only for intermediate postsynaptic depolarizations when the
voltage-dependent~Mg$^{2+}$ block of these receptors is released
\citep{JahrStevens1990}. In this fashion, the possible effect of~SICs
on~STDP would be confined in a time window around $\Delta t \ge 0$ for
which coincidence with pre- and postsynaptic spikes allows for robust
depolarization of postsynaptic spines. Outside this window
instead,~SICs would be negligible, and plasticity induction would
essentially depend on mere pre- and postsynaptic spiking rescuing the experimental observation of no synaptic modification for large spike timing intervals
\citep{SjostromNelson_CON2002,Caporale_ARN2008}.

On the other hand, even without considering voltage-dependence
of~SIC-mediating NMDARs, the precise timing of~SICs with respect to
pre/post pairs, is predicted by our analysis, to be potentially
critical in determining strength and sign of plasticity. And similar
considerations could also hold for the onset time and duration of
modulations of synaptic release triggered by gliotransmission with
respect to the temporal features of plasticity-inducing stimuli
\citep{DePitta_FCN13}. This ultimately points to timing of glutamate
release by the astrocyte (and its downstream effects on synaptic
transmission) as a potential additional factor for associative
(Hebbian) learning, besides sole correlation between pre- and
postsynaptic activities \citep{Hebb1949,Gerstner_BC2002}. Remarkably,
this could also provide a framework to conciliate the possibility that
modest, sporadic~SICs that we predict would not substantially
affect~STDP (\figref{fig:stdp_sic}), could do so instead
\citep{Chen_PNAS2012}. Indeed our predictions are based on the average
number of~SICs within a given time window as documented in literature
rather than on the precise timing of those~SICs in that time
window. In this fashion for example, there is no distinction in terms
of effect on~STDP in our simulations, between a hypothetical scenario
of three~SICs randomly occurring on average every $\sim$10~s in a 30~s
time frame and the alternative scenario of three~SICs taking place within
the same time frame but in rapid succession
\citep[Figure~5B]{PereaAraque_JN2005}, as could happen following an
exocytic burst of glutamate release by the astrocyte
\citep{Marchaland_etal_JN2008,Santello_etal_Neuron2011,Sahlender_PTRSB2014}. Yet
the latter case could result in a dramatically different plasticity
outcome with respect to the former. While individual~SICs likely fail
to induce synaptic modification alone in fact, their occurrence in
rapid succession would instead allow postsynaptic~Ca$^{2+}$ levels to
quickly increase beyond one of the thresholds for plasticity
induction. Furthermore, this increase could further be boosted by
coincidence of~SICs with pre- and postsynpatic activity, ultimately
accounting for robust~LTP, as indeed predicted by other theoretical
investigations \citep{Wade_PLoSOne2011}. However, to complicate this intriguing
scenario is the observation that glutamatergic
gliotransmission \citep{Santello_etal_Neuron2011}, and even more so
astrocyte-mediated~SICs \citep{ParriCrunelli2001,Bardoni_JP2010}, are
likely not deterministic but rather stochastic processes. Therefore,
it would ultimately be interesting to understand how this
stochasticity could affect neuronal activity and shape learning
\citep{Porto_PLoSOne2011}.

To conclude, our analysis provides theoretical arguments in support of the hypothesis that,
beyond neuronal firing, astrocytic gliotransmission could represent an
additional factor in the regulation of activity-dependent plasticity
and learning
\citep{BainsOliet_TINS2007,Min_FCN2012,DePitta_Neurosci2015}. This
could occur in a variegated fashion by both presynaptic and
postsynaptic elements targeted by glutamatergic gliotransmission, with
possibly diverse functional consequences. Nonetheless, the practical
observation in future experiments of a possible mechanism of action of
glutamatergic gliotransmission on activity-dependent plasticity will
depend on the implementation of novel specific plasticity-inducing
protocols that match possible stringent temporal and spatial
dynamical constraints defining the complex nature of neuron-astrocyte
interactions.

\section*{Acknowledgements}

M.D.P. wishes to thank Hugues Berry for insightful discussions, Marcel
Stimberg, Romain Brette and members of Brette's group at the Institut
de la Vision (Paris) for hospitality and assistance in implementing
simulations of this work in Brian 2.0. This work was supported by a FP7~Maria Sk\l odowska-Curie International Outgoing Fellowship to M.D.P. (Project~331486 ``Neuron-Astro-Nets''). The authors declare no
conflicts of interests.

\begin{appendices}
  \renewcommand\thetable{\thesection\arabic{table}}
  \renewcommand\thefigure{\thesection\arabic{figure}}
  \section{Modeling methods} \label{app:model}
  \subsection{Synapse model with glutamatergic gliotransmission}\label{app-sec:ca-model}
  \subsubsection{Synaptic release}
To study modulation of short-term synaptic plasticity by gliotransmitter-bound extrasynaptically-located presynaptic receptors we extend the model originally introduced by \citet{DePitta_PCB11} for astrocyte-mediated heterosynaptic modulation of synaptic release to also account for the homosynaptic scenario. For the sake of clarity, in the following we will limit our description to excitatory (glutamatergic) synapses. Accordingly, synaptic glutamate release is described following \citet{Tsodyks_LesHouches2005}, whereby upon arrival of an action potential~(AP) at time $t_k$, the probability of glutamate resources to be available for release ($u_S$) increases by a factor $u_0$, while the readily-releasable glutamate resources ($x_S$) decrease by a fraction $r_S(t_k)=u_S(t_k^+)x_S(t_k^-)$, corresponding to the fraction of effectively released glutamate. In between APs, glutamate resources are reintegrated at rate $1/\tau_d$ while $u_S$ decays to zero at rate $1/\tau_f$. The equations for $u_S,\,x_S$ thus read \citep{Tsodyks_LesHouches2005}
\begin{align}
\tau_f\dd{t}u_S &= -u_S + \sum_{k}u_{0}(1-u_S)\,\delta(t-t_k)\tau_f\label{eq:us}\\
\tau_d\dd{t}x_S &= 1-x_S -\sum_{k}r_S(t)\,\delta(t-t_k)\tau_d\label{eq:xs}
\end{align}
The parameter $u_0$ in the above equations may be interpreted as the synaptic release probability at rest. Indeed, when the period of incoming APs is much larger than the synaptic time scales $\tau_d,\,\tau_f$, in between APs $u_S\rightarrow 0,\,x_S\rightarrow 1$ -- that is the synapse is ``at rest" --, while, upon arrival of an AP, the probability of glutamate release from the synapse equals $u_0$.
\subsubsection{Neurotransmitter time course}
Assuming a total vesicular glutamate concentration of $Y_T$, the released glutamate, expressed as concentration in the synaptic cleft is then equal to $Y_{rel}(t_k)=\varrho_c\,Y_T\,r_S(t_k)$, where $\varrho_c$ represents the ratio between vesicular and synaptic cleft volumes. The time course of synaptically-released glutamate in the cleft ($Y_S$) depends on several mechanisms, including clearance by diffusion, uptake and/or degradation \citep{Clements1996,Diamond_JN2005}. In the simplest approximation, the contribution of these mechanisms to glutamate time course in the cleft may be modeled by a first order degradation reaction of characteristic time $\tau_c$ \citep{DestexheJCompNeurosci1994} so that
\begin{align}
\tau_c\dd{t}Y_S &= -Y_S + \sum_{k}Y_{rel}\,\delta(t-t_k)\tau_c\label{eq:ys}
\end{align}
\subsubsection{Astrocytic calcium dynamics}
We assume that only a fraction $\zeta$ of released glutamate binds to postsynaptic receptors, while the remainder $1-\zeta$ fraction spills out of the cleft and activates astrocytic metabotropic receptors which trigger astrocytic~Ca$^{2+}$ signaling. The latter is modeled following \citet{Wallach_PCB2014} and results from the interplay of four quantities: (i)~the fraction of activated astrocytic receptors ($\gamma_A$); (ii)~the cytosolic~IP$_3$ ($I$) and (iii)~Ca$^{2+}$ concentrations ($C$) in the astrocyte; and (iv)~the fraction of deinactivated~IP$_3$ receptors (IP$_3$Rs) on the membrane of the astrocyte's endoplasmic reticulum~(ER) that mediate~Ca$^{2+}$-induced~Ca$^{2+}$-released from this latter ($h$). In particular, considering each quantity separately, the fraction of astrocytic receptors bound by synaptic glutamate may be approximated, at first instance, by a first order binding reaction and thus evolves according to \citep{Wallach_PCB2014}
\begin{align}
\tau_A \dd{t}\gamma_A &= -\gamma_A + O_M (1-\zeta) Y_S (1-\gamma_A)\tau_A \label{eq:gammaa}  
\end{align}
with $\tau_A$ representing the characteristic receptor deactivation (unbinding) time constant. Cytosolic~IP$_3$ results instead from the complex~Ca$^{2+}$-modulated interplay of phospholipase C$\beta$- and C$\delta$-mediated production and degradation by~IP$_3$ 3-kinase (3K) and inositol polyphosphatase 5-phosphatase (5P) \citep{ZhangMuallem1993,SimsAllbritton1998,RebecchiPentyala2000,Berridge_etal_NatRev2003}, and evolves according to the mass balance equation \citep{DePitta_JOBP2009}
\begin{align}
\dd{t}I     &= J_\beta(\gamma_A) + J_\delta(C,I) - J_{3K}(C,I) - J_{5P}(I) \label{eq:I}\\
\end{align}
where
\begin{flalign*}
J_\beta(\gamma_A) &= O_\beta \, \gamma_A
& J_\delta(C,I)   &= O_\delta \, \frac{\kappa_\delta}{\kappa_\delta + I}\Hill{C^2}{K_\delta}\\ 
J_{3K}(C)         &= O_{3K}\, \Hill{C^4}{K_D}\Hill{I}{K_3}
& J_{5P}(I)       &= \Omega_{5P}\, I\\
\end{flalign*}
and \Hill{x^n}{K} denotes the sigmoid (Hill) function $x^n/(x^n + K^n)$. Finally, cytosolic~Ca$^{2+}$ and the~IP$_3$R gating are described by a set of Hodgkin-Huxley-like equations according to the model originally introduced by \citet{LiRinzel1994}:
\begin{align}
\dd{t}C     &= J_C(C,h,I) + J_L(C) - J_P(C) \label{eq:C}\\
\dd{t}h     &= \frac{h_\infty(C,I)-h}{\tau_h(C,I)} \label{eq:h}
\end{align}
where $J_C,\,J_L,\,J_P$ respectively denote the~IP$_3$R-mediated~Ca$^{2+}$-induced~Ca$^{2+}$-release from the ER ($J_C$), the~Ca$^{2+}$ leak from the ER ($J_L$), and the~Ca$^{2+}$ uptake from the cytosol back to the~ER by serca-ER~Ca$^{2+}$/ATPase pumps ($J_P$) \citep{DePitta_JOBP2009}. These terms, together with the~IP$_3$R deinactivation time constant ($\tau_h$) and steady-state propability ($h_\infty$), are given by \citep{LiRinzel1994,DePitta_CognProc2009}
\begin{flalign*}
J_{C}(C,h,I)    &= \Omega_C \, m_\infty^3 h^3 \, (C_T-(1+\varrho_A)C) 
& m_\infty(C,I) &= \Hill{I}{d_{1}}\Hill{C}{d_{5}}\\
J_L(C)          &= \Omega_L \, (C_T-(1+\varrho_A)C) 
& J_{P}(C)      &= O_P \Hill{C^2}{K_P}\\
h_{\infty}(C,I) &= d_{2}\frac{I + d_{1}}{d_2(I+d_1)+(I+d_3)C} 
& \tau_h(C,I)   &= \frac{I+d_3}{\Omega_{2}(I+d_1)+O_2(I+d_3)C}
\end{flalign*}
A detailed explanation of the parameters of the astrocyte model may be found in the Table in \appref{TA:All-Parameters}.
\subsubsection{Calcium-dependent glutamatergic gliotransmission}\label{app-sec:glt-rel}
Astrocytic glutamate exocytosis is modeled akin to synaptic glutamate release, assuming that a fraction $x_A(t)$ of gliotransmitter resources is available for release at any time $t$. Then, every time $t_j$ that astrocytic~Ca$^{2+}$ increases beyond a threshold concentration $C_\theta$, a fraction of readily-releasable astrocytic glutamate resources, i.e. $r_A(t_j) = U_A\,x_A(t_j^-)$, is released into the periastrocytic space, and later reintegrated at rate $1/\tau_A$. Hence, $x_A(t)$ evolves according to \citep{DePitta_PCB11}
\begin{align}
\tau_G\dd{t}x_A &= 1-x_A -\sum_{j}r_A(t)\,\delta(t-t_j)\tau_G\label{eq:xa}
\end{align}
Similarly to \eqref{eq:ys}, we may estimate the contribution to glutamate concentration in the periastrocytic space ($G_A$), resulting from a quantal glutamate release event by the astrocyte at $t=t_j$, as $G_{rel}(t_j)=\varrho_e\, G_T\, r_A(t_j)$, where $G_T$ represents the total vesicular glutamate concentration in the astrocyte, and $\rho_e$ is the volume ratio between glutamate-containing astrocytic vesicles and periastrocytic space. Then, assuming a clearance rate of glutamate of $1/\tau_e$, the time course of astrocyte-derived glutamate in the extracellular space comprised between the astrocyte and the surrounding synaptic terminals is given by
\begin{align}
\tau_e\dd{t}G_A &= -G_A + \sum_{j}G_{rel}(t)\,\delta(t-t_j)\tau_e\label{eq:ga}
\end{align}
\subsubsection{Presynaptic pathway for glutamatergic gliotransmission}
The extracellular glutamate concentration sets the fraction of bound extrasynaptically-located presynaptic receptors ($\gamma_S$) according to \citep{DePitta_PCB11}
\begin{align}
\tau_P \dd{t}\gamma_S = -\gamma_S + O_P\,(1-\gamma_S)G_A\tau_P \label{eq:gammas}
\end{align}
where $O_P$ and $\tau_P$ respectively denote the rise rate and the decay time of the effect of gliotransmission on synaptic glutamate release. It is then assumed that modulations of synaptic release by gliotransmitter-bound presynaptic receptors are brought forth by modulations of the resting synaptic release probability, i.e. $u_0=u_0(\gamma_S)$. In an attempt to consider a mechanism as general as possible, rather than focusing on a specific functional dependence for $u_0(\gamma_S)$, we consider only the first order expansion of this latter \citep{DePitta_PCB11}, that is
\begin{align}
u_0(\gamma_S) \approx U_0 + (\xi-U_0)\gamma_S \label{eq:u0}
\end{align}   
where $U_0$ denotes the synaptic release probability at rest in the absence of gliotransmission, whereas the parameter $\xi$ lumps, in a phenomenological way, the information on the effect of gliotransmission on synaptic release. For $0 \le \xi < U_0$, $u_0$ decreases with $\gamma_S$, consistent with a ``release-decreasing" effect of astrocytic glutamate on synaptic release. This could be the case, for example, of astrocytic glutamate binding to presynaptic kinate receptors or group~II/III mGluRs \citep{AraqueParpuraEJN1998a,LiuNedergaard2004,Liu_PNAS2004}. Vice versa, for $U_0 < \xi \le 1$, $u_0$ increases with $\gamma_S$, consistent with a ``release-increasing" effect of astrocytic gliotransmitter on synaptic release, as in the case of presynaptic NMDARs or group I mGluRs \citep{FiaccoMcCarthy2004,JourdainVolterra2007,PereaAraque_Science2007,Navarrete_Neuron2010,Bonansco_etal_EJN2011,Navarrete_PB2012,Perea_NatComm2014}.   Finally, the special case where $\xi = U_0$ corresponds to occlusion, that is coexistence and balance between release-decreasing and release-increasing glutamatergic gliotransmission at the same synapse resulting in no net effect of this latter on synaptic release.\\
\indent
Although based on glutamatergic synapses, the set of equations~\ref{eq:us}--\ref{eq:u0} provides a general description for modulations of synaptic release mediated by glutamatergic gliotransmission that could also be easily extended to other types of excitatory synapses \citep{Pankratov_JGP_2007} as well as inhibitory synapses \citep{KangNedergaard1998,Serrano_etal_JN2006,Liu_PNAS2004,Losi_PTRSB2014}.
\subsubsection{Postsynaptic pathways for glutamatergic gliotransmission: slow inward currents}
Postsynaptic astrocyte-mediated slow inward currents take place through extrasynaptic NMDA receptors. The subunit composition of these receptors however remains unclear \citep{Papouin_PTRSB2014}. Several studies reported SICs to be inhibited by antagonists of NR2B-containing NMDARs \citep{FellinCarmignoto_Neuron2004,Shigetomi_JN2008,Pirttimaki_JN2011}, there is also evidence that other NMDAR types could be involved possibly  subunits could be involved such as NR2C or NR2D \citep{Bardoni_JP2010}. Being mediated by NMDA receptors, SICs are likely affected by voltage-dependence of the Mg$^{2+}$ block of these receptors. Although there is evidence that SICs rate and frequency could indeed depend on extracellular Mg$^{2+}$ \citep{FellinCarmignoto_Neuron2004}, the effective nature of the possible voltage dependence of SICs has not been elucidated. Moreover, the potential diversity of subunit composition of receptors mediated SICs could also result in different voltage dependencies, strong for NR2B-containing receptors, akin to postsynaptic NMDARs, and weak otherwise \citep{Cull-Candy_CON2001}. Accordingly, in this study we neglect the possible voltage-dependence of SICs arguing that this is not substantially changing the essence of our results (see Discussion). In this fashion, denoting postsynaptic SICs by $i_A(t)$, we model them by a difference of exponentials according to 
\begin{align}\label{eq:sic}
\tau_S^r\,\dd{t}i_A(t) &= -i_A(t) + \hat{I}_A B_A(t)\,\tau_S^r\\
\tau_S\,\dd{t}B_A(t) &= -B_A(t) + \hat{J}_A G_A(t)\,\tau_S
\end{align}
where $\tau_S^r,\,\tau_S$ respectively are rise and decay time constants for~SICs. The two scaling factors $\hat{I}_A,\,\hat{J}_A$ are taken such that the SIC maximum in correspondence with an event of glutamate release by the astrocyte is equal to a constant value $I_A$ (see \appref{app:par-estimation}).
\subsubsection{Postsynaptic neuron}
Postsynaptic action potential firing is modeled by a leaky integrate-and-fire (LIF) neuron \citep{Burkitt_BC2006,Fourcaud_NC2002}. Accordingly, the membrane potential ($v$) of the postsynaptic neuron evolves as
\begin{align}
\tau_m\dd{t}v &= E_L-v + i_S(t) + i_A(t)
\end{align}
where $\tau_m$ denotes the membrane time constant and $i_S(t)$ represents the excitatory synaptic input to the neuron. Every time $v$ crosses the firing threshold $v_\theta$, an AP is emitted and the membrane potential is reset to $v_r$ and kept at this value for the duration of a refractory period $\tau_r$.\\
\indent
For synaptic currents, we only consider the AMPA receptor-mediated fast component of EPSCs. Accordingly, we consider two possible descriptions for $i_S(t)$. In Figures~\ref{fig:pre-loop} and~\ref{fig:syn_filtering}, we assume that the rise time of synaptic currents is very short compared to the relaxation time of these latter \citep{Spruston_JP1995,MageeCook_NatureNeurosci2000,AndrasfalvyMagee_JN2001}, so that $i_S(t)$ can be modeled by a sum of instantaneous jumps of amplitude $\hat{J}_S$, each followed by an exponential decay \citep{Fourcaud_NC2002}, i.e.
\begin{align}\label{eq:psc-1}
\tau_N\dd{t}i_S(t) &= -i_S(t) + \hat{J}_S \zeta Y_S(t)\,\tau_N
\end{align}
In the presence of gliotransmitter-mediated slow-inward currents instead (\figref{fig:sics}), we model synaptic currents similarly to these latter, that is
\begin{align}\label{eq:psc}
\tau_N^r\dd{t}i_S(t) &= -i_S(t) + \hat{I}_S B_S(t)\,\tau_N^r\\
\tau_N\dd{t}B_S(t) &= -B_S(t) + \hat{J}_S \zeta Y_S(t)\,\tau_N
\end{align}
where $\tau_N^r,\,\tau_N$ respectively denote EPSC rise and decay time constants; the scaling factor $\hat{J}_S$ is taken such that synaptic releases result in unitary increases of the gating variable $B_S$, and similarly, $\hat{I}_S$ is set such that an increases in synaptic current ensuing from quantal synaptic release equals to $I_S$.

\subsection{Spike-timing--dependent plasticity}
\subsubsection{Postsynaptic calcium dynamics}
Spike-timing--dependent plasticity is modeled by the nonlinear calcium model introduced by \citet{Graupner_PNAS2012}, which was modified to include short-term synaptic plasticity as well as astrocyte-mediated SIC contribution to postsynaptic~Ca$^{2+}$. Accordingly, postsynaptic~Ca$^{2+}$ dynamics, $c(t)$, results from the sum of three contributions: (i)~Ca$^{2+}$ transients mediated by NMDA receptors, activated by synaptic glutamate whose release probability from the presynaptic bouton may be modulated by gliotransmission ($c_{pre}$), (ii)~Ca$^{2+}$ transients mediated by gliotransmitter-triggered NMDA-mediated~SICs ($c_{sic}$), (iii)~Ca$^{2+}$ transients due to activation of voltage-dependent~Ca$^{2+}$ channels (VDCCs) by postsynaptic backpropagating APs ($c_{post}$). All three~Ca$^{2+}$ transients are accounted for by a difference of exponentials. In particular, presynaptic~Ca$^{2+}$ transients are described by
\begin{align}
\tau_{pre}^r\dd{t}c_{pre} &= -c_{pre} + \hat{C}_{pre}R_{pre}\tau_{pre}^r\\
\tau_{pre}\dd{t}R_{pre} &= -R_{pre} + W_N\,\zeta\,Y_S\,\tau_{pre}
\end{align}
where $\tau_{pre}^r,\,\tau_{pre}$ are rise and decay time constants of the~Ca$^{2+}$ transient; $\hat{C}_{pre}$ is a normalization constant such that the maximal amplitude of the transient is $C_{pre}$; $W_N$ is the ``weight" of presynaptic~Ca$^{2+}$ transients triggered by synaptic glutamate ($Y_S$, \eqref{eq:ys}).\\
\indent
Similarly to presynaptic ones, calcium transients due to by SICs mediated by gliotransmission are given by
\begin{align}
\tau_{sic}^r\dd{t}c_{sic} &= -c_{sic} + \hat{C}_{sic}R_{sic}\tau_{sic}^r\\
\tau_{sic}\dd{t}R_{sic} &= -R_{sic} + W_A\,G_A\,\tau_{sic}
\end{align}
where $\tau_{sic}^r,\,\tau_{sic}^d$ are the rise and decay time constants of the~Ca$^{2+}$ transient; and $\hat{C}_{sic}$ is a normalization constant such that the maximal amplitude of the transient is $C_{sic}$; $W_A$ is the ``weight" of~SIC-mediated~Ca$^{2+}$ transients triggered by perisynaptic gliotransmitter ($G_A$, \eqref{eq:ga}).\\
\indent
Finally, postsynaptic~Ca$^{2+}$ transients caused by bAPs are modeled by \citep{Graupner_PNAS2012}
\begin{align}
\tau_{post}^r\dd{t}c_{post} &= -c_{post} + \hat{C}_{post}R_{post}\tau_{post}^r\\
\tau_{post}\dd{t}R_{post} &= -R_{post} + (1+\eta c_{pre})\sum_i \delta(t-t_i)\tau_{post}
\end{align}
where the sum goes over all postsynaptic APs occurring at times $t_i$; $\tau_{post}^r,\,\tau_{post}$ are the rise and decay time constants of the~Ca$^{2+}$ transient; $\hat{C}_{post}$ is a scaling factor such that the maximal amplitude of the transient is $C_{post}$, and the parameter $\eta$ implements by which amount the bAP-evoked~Ca$^{2+}$ transient is increased in case of preceding presynaptic stimulation.
\subsubsection{Synaptic efficacy}
The temporal evolution of synaptic efficacy $\rho(t)$ depends on postsynaptic~Ca$^{2+}$ dynamics $c(t)=c_{pre}(t)+c_{sic}(t)+c_{post}(t)$ and is described by the first-order differential equation \citep{Graupner_PNAS2012}
\begin{align}
\tau_\rho \dd{t}\rho = -\rho(1-\rho)(\rho_\star-\rho) + \gamma_p(1-\rho)\functionof{\Theta}{c(t)-\theta_p} -\gamma_d\,\rho\,\functionof{\Theta}{c(t)-\theta_d} + \mathrm{Noise}(t)
\end{align}
where $\rho_\star$ is the boundary of the basins of attraction of UP and DOWN states of synaptic efficacy, that is the states for which $\rho=1$ and $\rho=0$ respectively; $\theta_d,\,\theta_p$ denote the depression (LTD) and potentiation (LTP) thresholds, and $\gamma_d,\,\gamma_p$ measure the corresponding rates of synaptic decrease and increase when these thresholds are exceeded; $\Theta (\cdot)$ denotes the Heaviside function, i.e. $\functionof{\Theta}{c-\theta}=0$ for $c<\theta$ and $\functionof{\Theta}{c-\theta}=1$ for $c\ge \theta$. The last term lumps an activity-dependent noise term in the form of $\mathrm{Noise}(t)=\sigma \sqrt{\tau_\varrho} \sqrt{\functionof{\Theta}{c(t)-\theta_d}+\functionof{\Theta}{c(t)-\theta_p}}\cdot \varpi(t)$ where $\varpi(t)$ is a Gaussian white noise process with unit variance density. This term accounts for activity-dependent fluctuations stemming from stochastic neurotransmitter release, stochastic channel opening and diffusion \citep{Graupner_PNAS2012}.

\subsection{STDP curves}\label{app-sec:synaptic-strength}
To construct the~STDP curves of Figures~\ref{fig:stdp-curves} and~\ref{fig:stdp_sic}, we follow the rationale originally described by \citet{Graupner_PNAS2012}, and consider the average synaptic strength of a synaptic population after a stimulation protocol of~$n$ pre-post (or post-pre) pairs presented at time interval~$T$. With this aim, synaptic strength is surmised to be linearly related to $\rho$ as $w=w_0+\rho(w_1-w_0)$, where $w_0,\,w_1$ are the synaptic strength of the~DOWN/UP states for which $\rho_0$, i.e. the initial value of $\rho$ at $t=0$, is~0 or~1 respectively. In this fashion, $w$ may be thought as a rescaled version of equivalent experimental measures of synaptic strength such as the excitatory postsynaptic potential~(EPSP) or current~(EPSC) amplitude, the initial~EPSP slope or the current in a 2-ms windows at the EPSC peak. Accordingly, before a stimulus protocol, a fraction $\beta$ of the synapses is taken in the DOWN state, so that the average initial synaptic strength is $\bar{w}_0=\beta\,w_0 + (1-\beta)w_1$. Then, after the stimulation protocol, the ensuing average synaptic strength is $\bar{w}_1 = w_0((1-\mathcal{U})\beta + \mathcal{D}(1-\beta)) + w_1(\mathcal{U}\beta + (1-\mathcal{D})(1-\beta))$ where $\mathcal{U},\,\mathcal{D}$ represent the~UP and~DOWN transition probabilities respectively. As in experiments, we consider the change in synaptic strength as the ratio between the average synaptic strengths after and before the stimulation, i.e.
\begin{align}
\Delta \varrho &= \frac{(1-\mathcal{U})\beta + \mathcal{D}(1-\beta) + b(\mathcal{U}\beta+(1-\mathcal{D})(1-\beta))}{\beta+(1-\beta)b} 
\end{align}
with $b=w_1/w_0$. Under the hypotheses of this study, that $T\ll \tau_\varrho$ and $\gamma_d,\,\gamma_p$ are large, the transition probabilities $\mathcal{U},\,\mathcal{D}$ may be analytically solved and read \citep{Graupner_PNAS2012}
\begin{align}
\mathcal{U}(\rho_0) &= \frac{1}{2}\left(1+\erf \left(-\frac{\rho_\star-\bar{\rho}+(\rho-\rho_0)\expp{-\frac{nT}{\tau}}}{\sqrt{\sigma_\rho}\left(1-\expp{-\frac{2nT}{\tau}}\right)} \right) \right)\\
\mathcal{D}(\rho_0) &= \frac{1}{2}\left(1-\erf \left(-\frac{\rho_\star-\bar{\rho}+(\rho-\rho_0)\expp{-\frac{nT}{\tau}}}{\sqrt{\sigma_\rho}\left(1-\expp{-\frac{2nT}{\tau}}\right)} \right) \right)
\end{align}  
where $\erf$ denotes the standard Error Function, defined as $\erf(x)=\frac{2}{\sqrt{\pi}}\int_0^x \expp{-t^2}\mathrm{d}t$ and
\begin{align*}
\bar{\rho} &= \frac{\Gamma_p}{\Gamma_d + \Gamma_p},
& \sigma_\rho^2 &= \frac{\alpha_d+\alpha_p}{\Gamma_d + \Gamma_p}\sigma^2,
& \tau = \frac{\tau_\rho}{\Gamma_d + \Gamma_p}
\end{align*}
with $\Gamma_i = \gamma_i\,\alpha_i$ and $\alpha_i=\frac{1}{nT}\int_0^{nT}\functionof{\Theta}{c(t)-\theta_i}\mathrm{d}t$ with $i=d,\,p$.

\subsection{Simulations}
The model was implemented in Brian~2.0 \citep{Stimberg_FNI2014}. Simulations and data analysis were serendipitously designed and performed by the open source programming language Python~2.7 \citep{Python_2_7}. The code is available online at $<$\textit{add url}$>$.
\newpage
\section{Parameter estimation}\label{app:par-estimation}
\subsection{Synaptic parameters}
Glutamate release probability~$U_{0}$ of central excitatory synapses is generally comprised between~$\sim$0.09 \citep{Schikorski1997} and~$\sim$0.6--0.9 \citep{StevensWang_Neuron1995,Markram_etal_Neuropharmacol1998}, with lower values mostly consistent with facilitating synapses \citep{Murthy_etal_Neuron1997}. Facilitation time constants~$\tau_{f}$ may be estimated by the decay time of intracellular~Ca$^{2+}$ increases at presynaptic terminals upon arrival of action potentials \citep{Regehr_etal_JN1994,Emptage_etal_Neuron2001}. With this regard, typical decay times for~Ca$^{2+}$ transients are reported to be~$<$500~ms \citep{Emptage_etal_Neuron2001}, with an upper bound between~0.65--2~s \citep{Regehr_etal_JN1994}. Concerning depression time constants instead, experiments have reported glutamate-containing vesicles in the readily releasable pool to preferentially undergo rapid endocytosis within~1--2~s after release \citep{Pyle_etal_Neuron2000}, although vesicle recycling could also be as fast as~10--20~ms \citep{StevensWang_Neuron1995,BrodyYue_JN2000}.

Estimates in hippocampal synapses suggest that the readily releasable pool could count between~2 and~27 vesicles \citep{Schikorski1997} which are essentially spherical with average outer diameter~$d_{S}$ equal to~39.2$\pm$11.4~nm and in the range of~23--49~nm \citep{HarrisSultan1995,Bergersen_CC2011}. Subtracting to this value a~6~nm-thick vesicular membrane \citep{Schikorski1997}, the inner diameter of a vesicle can then be estimated between~16--38~nm, corresponding to a mean vesicular volume~$\Lambda_{S}$ in the range of~$2.1\cdot 10^{-21}$--28.7$\cdot 10^{-21}$~dm$^{3}$. Given that vesicular glutamate concentration is reported in the range of~60--210~mM \citep{HarrisSultan1995,Danbolt_ProgNeurobiol2001,Bergersen_CC2011}, then considering a pool of~10~vesicles with average diameter of~30~nm~(i.e.~average volume~$\Lambda_{S} \approx 14.1\cdot 10^{-21}$~dm$^{3}$) and average vesicular neurotransmitter concentration of~60--100~mM \citep{Bergersen_CC2011}, the total neurotransmitter vesicular content ranges up to~$Y_{T}$~=~(10)(60--100~mM)~=~300--1000~mM. Assuming a typical neurotransmitter release time of~$t_{rel}=25$~$\mu$s \citep{RaghavacariLisman2004} and a diffusion constant for glutamate in the synaptic cleft of~$D_{\textrm{Glu}}=0.33$~$\mu$m$^{2}$/ms \citep{Nielsen_etal_Neuron2004}, the average diffusion length~($\ell _{c}$) of a glutamate molecule from the release site can be estimated by the Einstein-Smoluchowski relationship \citep{Schikorski1997} whereby~$\ell_{c} = \surd (2 \cdot D_{\textrm{Glu}} \cdot t_{rel})=\surd(2\cdot$0.33~$\mu$m$^{2}$/ms$\,\cdot \,$0.025~ms$)\approx$~0.129~$\mu$m.
Thus, the associated mixing volume~$\Lambda_{c}$, namely the effective diffusion volume which the released glutamate has rapid access to, can be estimated by the volume of the disk of radius~$\ell_{c}$ and thickness~$h_{c}$ equal to the average width of the synaptic cleft \citep{Barbour_JN2001}. Considering~$h_{c} = 20$~nm \citep{Schikorski1997}, it is:~$\Lambda_{c} = \pi\, \ell_{c}^{2} h_{c} = \pi\cdot$(0.129~$\mu$m)$^{2}$(0.020~$\mu$m)~$\approx 8.89\cdot 10^{-18}$~dm$^{3}$ which falls in the experimental range of volumes of nonsynaptic interfaces at hippocampal synapses elsewhere reported \citep{VenturaHarris1999}. Considering vesicular release from at least~3 independent sites \citep{Oertner_NN2002}, it follows that the ratio between vesicle volume and mixing volume is~$\varrho_{c}  = \Lambda_{S}/ \Lambda_{c}$=(3)($14.1\cdot 10^{-21}$~dm$^{3}$)/($8.89\cdot 10^{-18}$~dm$^{3}$)~$\approx 0.005$, so that the contribution to the concentration of glutamate in the extracellular space following a release event, is~$Y_{rel} = \rho_{c}\cdot U_{0} \cdot Y_{T}$. Hence, for a sample value of~$U_{0}=0.5$ \citep{StevensWang_Neuron1995} with a choice of~$Y_{T}=500$~mM for example, the latter equals to~$Y_{rel}=(0.005)(0.5)(500$~mM)~$\approx$1.25~mM.\\
\indent
Such released glutamate is then rapidly cleared from the extracellular space by combined action of diffusion and uptake by transporters \citep{BarbourHausser_TiNS1997}. As a result the time course of glutamate in the synaptic cleft is short, with an estimated decay constant between~$\sim$2--10~ms \citep{Clements1992,Diamond_JN2005}. However, slower clearance times could also be possible since resting glutamate concentrations in the extracellular space surrounding activated synapses are recovered only~$\sim$100~ms after the stimulus \citep{HermanJahr_JN2007,Okubo_etal_PNAS2010}. Based on these considerations, we consider an intermediate value of glutamate clearance time of $\tau_c = 25$~ms.
\subsection{Astrocyte parameters}
Astrocyte parameters reported in \appref{TA:All-Parameters} were estimated on extensive numerical explorations of the astrocyte model aimed at reproducing experimental whole-cell~Ca$^{2+}$ elevations with rise and decay time constants respectively in the ranges of~3--20~s and~3--25~s and with full-width half-maximum~(FWHM) values between~5--160~s \citep{HiraseBuzsaki2004,NimmerjahnHelmchen2004,WangNedergaard2006}. In doing so, we considered a ratio between~ER and cytoplasmic volumes~($\varrho_{A}$)~of~0.18 in line with the experimental observation that the probability of~ER localization in the cytoplasmic space at astrocytic somata is between~$\sim$20--70\% \citep{Pivneva_CellCalcium2008}. Moreover, the cell's total~Ca$^{2+}$ concentration (measured with respect to the cytoplasmic volume) was fixed at $C_T=2$~$\mu$M, while~Ca$^{2+}$ affinity of sarco-ER pumps (SERCAs) was taken equal to 0.05~$\mu$M \citep{Lytton1992,VandecaetsbeekVangheluwe_PNAS2009}, assuring peak~Ca$^{2+}$ concentrations~$<$5~$\mu$M in agreement with experiments \citep{ParpuraHaydon2000,KangOthmer_Chaos2009}.\\
\indent
Activation rate and unbinding time constants~$O_{A},\,\tau_{A}$ of astrocytic receptors may be estimated by rise times of agonist-triggered~Ca$^{2+}$ signals. With this regard, application of~50~$\mu$M of~DHPG, a potent agonist of group~I subtype~5~mGluRs -- the main type of~mGluRs expressed by astrocytes \citep{AronicaTroost2003} --, triggered submembrane~Ca$^{2+}$ signals characterized by a rise time~$\tau_{r}=0.272\pm 0.095$~s. Because~mGluR5 affinity~($K_{0.5}$) for~DHPG is~$\sim$2~$\mu$M \citep{Brabet_NP1995}, that is much smaller than the applied agonist concentration, receptor saturation may be assumed in those experiments so that the receptor activation rate by~DHPG~($O_{\textrm{DHPG}}$) can be expressed as a function of~$\tau_{r}$ \citep{Barbour_JN2001} whereby~$O_{\textrm{DHPG}}\approx \tau_{r}(50$~$\mu$M$)^{-1}=0.055$--0.113~$\mu$M$^{-1}$s$^{-1}$ and, accordingly,~$\Omega_{\textrm{DHPG}}=1/O_{\textrm{DHPG}}K_{0.5}\approx 4$--10~s. Corresponding rate constants for glutamate may then be estimated by the latter, assuming similar kinetics yet with~$K_{0.5}=K_{A}=1/O_{A}\tau_A \approx 3$--10~$\mu$M \citep{Daggett_NP1995}, that is~1.5--5-fold larger than~$K_{0.5}$ for~DHPG. Moreover, since rise times of~Ca$^{2+}$ signals triggered by non-saturating physiological stimuli are somehow faster than in the case of~DHPG \citep{Panatier_etal_Cell2011}, it may be assumed that~$O_{N}>O_{\textrm{DHPG}}$. With this regard, for a choice of~$O_{A}\approx 3\,O_{\textrm{DHPG}}= 0.3$~$\mu$M$^{-1}$s$^{-1}$, with~$K_{A}=6$~$\mu$M such that~$\tau_A = 1/ ($0.3~$\mu$M$^{-1}$s$^{-1}$)/(6~$\mu$M$)=0.55$~s, a peak synaptic glutamate concentration of~$Y_{rel}=1200$~$\mu$M, with~$\tau_c=25$~ms results in a maximum average fraction of bound receptors of~$\sim$0.75--0.9 that occurs within~$\approx 70$~ms from synaptic release, in good agreement with experimentally-reported rise times.
\subsection{Gliotransmission}
Exocytosis of glutamate from astrocytes is reported to occur by Ca$^{2+}$ concentrations increasing beyond a threshold value $C_\theta \approx 0.15-0.8$~$\mu$M. In this study we specifically consider $C_\theta = 0.5$~$\mu$M. Glutamate-containing vesicles found in astrocytic processes have regular~(spherical) shape with typical diameters~($d_{A}$) between~$\sim$20--110~nm \citep{BezziVolterra2004,Crippa_etal_JP2006}. The corresponding vesicular volume~$\Lambda_{A}$ then is between~$\sim$2--700$\cdot 10^{-21}$~dm$^{3}$. Vesicular glutamate content is approximately the same, or at least as low as one third of synaptic vesicles in adjacent nerve terminals \citep{MontanaParpura2006,JourdainVolterra2007,Bergersen_CC2011}. Thus, considering a range of synaptic vesicular glutamate content between~$\sim$60--150~mM \citep{Danbolt_ProgNeurobiol2001}, astrocytic vesicular glutamate concentration~($G_v$) is likely within~$\sim$20--150~mM \citep{Bergersen_CC2011}.\\
\indent
The majority of glutamate vesicles in astrocytic processes clusters in close proximity to the plasma membrane, i.e.~$<$100~nm, but about half of them is found within a distance of~40--60~nm from the ventral side of the membrane, suggesting existence of ``docked'' vesicles in astrocytic processes akin to synaptic terminals \citep{JourdainVolterra2007,Bergersen_CC2011}. Borrowing the synaptic rationale whereby docked vesicles approximately correspond to readily releasable ones \citep{Schikorski1997}, then the average number of astrocytic glutamate vesicles available for release~($n_{A}$) could be between~$\sim$1--6 \citep{JourdainVolterra2007}. Hence, the total vesicular glutamate releasable by an astrocyte may be estimated between~$G_{T}=n_{A}G_{A}=20$--900~mM.\\
\indent
Astrocytic vesicle recycling ($\tau_G$) likely depends on the mode of exocytosis. Both full-fusion of vesicles and kiss-and-run events have been observed in astrocytic processes \citep{BezziVolterra2004} with the latter seemingly occurring more often \citep{BezziVolterra2004,ChenZhou2005}. The fastest recycling pathway corresponds to kiss-and-run fusion, where the rate is mainly limited by vesicle fusion with plasma membrane and subsequent pore opening \citep{Valtorta_etal_TrendsCellBiol2001}. Indeed, reported pore-opening times in this case, can be as short as 2.0$\pm$0.3~ms \citep{ChenZhou2005}. The actual recycling time however, could be considerably longer if we take into account that, even for fast release events confined within~100~nm from the astrocyte plasma membrane, vesicle re-acidification could last~$\sim$1.5~s \citep{BowserKhakh2007}.\\
\indent
Considering a value for the diffusion constant of glutamate in the extracellular space of~$D_{\textrm{Glu}}=0.2$~$\mu$m$^{2}$/ms \citep{Nielsen_etal_Neuron2004}, and a vesicle release time $t_{rel}\approx 1$~ms \citep{ChenZhou2005}, the average diffusion distance travelled by astrocytic glutamate molecules into the extracellular space is~\citep{Schikorski1997} $\ell _{e}=\surd(2\cdot$0.2~$\mu$m$^{2}$/ms$\,\cdot \,$1~ms$)\approx$~0.63~$\mu$m. Then, assuming that the mixing volume of released glutamate~$\Lambda_{e}$ in the extracellular space coincides with one tenth (i.e.~0.1) of the ideal diffusion volume in free space \citep{RusakovKullmann_JN1998}, it is~$\Lambda_{e}=(0.1)\,4\,\pi\,\ell_{e}^{3}/3 \approx 10^{-16}$~dm$^{3}$. Accordingly, considering an average vesicular diameter~$d_{A}=50$~nm, so that~$\Lambda_{A}\approx 65\cdot 10^{-21}$~dm$^{3}$, the ratio~$\varrho_{e}$ between vesicular and mixing volumes for astrocytic glutamate diffusion can be estimated to be of the order of $\varrho_{e}=\Lambda_{A}/\Lambda_{e}=(65\cdot 10^{-21}$~dm$^{3})/10^{-16}$~dm$^{3}\approx 6.5\cdot 10^{-4}$. For an astrocytic pool of releasable glutamate of~$G_{T}=200$~mM, and a release probability~$U_{A}=0.6$, it follows that the extracellular peak glutamate concentration after exocytosis~is~$\hat{G}_{A}=(6.5\cdot 10^{-4})(0.6)(200$~mM$)\approx 78$~$\mu$M, in agreement with experimental measurements \citep{Innocenti_etal_JN2000}. Finally, imaging of extrasynaptic glutamate dynamics in hippocampal slices hints that glutamate clearance is fast and mainly carried out within~$<$300~ms of exocytosis \citep{Okubo_etal_PNAS2010}. Therefore, we consider a characteristic clearance time constant for glutamate in the periastrocytic space of $\tau_e=200$~ms.
\subsection{Presynaptic receptors}
We set the activation rate and inactivation time constants of presynaptic-receptors, i.e. $O_P$ and $\tau_P$, to reproduce the experimentally-reported rapid onset of the modulatory effect on synaptic release exerted by those receptors, namely within 1--5~s from glutamate exocytosis by the astrocyte, and the slow decay of this modulation, which is of the order of tens of seconds at least \citep{FiaccoMcCarthy2004,JourdainVolterra2007}. In particular, inhibition of synaptic release following activation of presynaptic~mGluRs by astrocytic glutamate could last from tens of seconds \citep{AraqueParpuraEJN1998a} to~$\sim$2--3~min \citep{LiuNedergaard2004}. Similarly, group~I~mGluR--mediated enhancement of synaptic release following a single~Ca$^{2+}$ elevation in an astrocyte process, may last as long as~$\sim$30--60~s \citep{FiaccoMcCarthy2004,PereaAraque_Science2007}. Values within~$\sim$1--2~min however, have also been reported in the case of an involvement of~NMDARs \citep{AraqueHaydon1998b,JourdainVolterra2007}. No specific assumption is made on the possible ensuing peak of receptor activation by a single glutamate release by the astrocyte.
\subsection{Postsynaptic neuron}
The membrane time constant of pyramidal neurons is typically between~20--70~ms \citep{PankratovKrishtal_JN2003,Routh_etal_JNeurophys2009} in correspondence of a membrane potential at rest in the range of $-66.5\pm$11.7~mV \citep{Spruston_JP1995,Magee_JN1998,MageeCook_NatureNeurosci2000,Otmakhova_etal_JN2002,GaspariniMiglioreMagee_JN2004,McDermott_etal_JP2006,Routh_etal_JNeurophys2009}. Firing threshold ($v_\theta$) values are on average around $-53\pm 2$~mV \citep{McDermott_etal_JN2003,GaspariniMiglioreMagee_JN2004,GaspariniMagee_JN2006}, ensuing in after-spike reset potentials ($v_r$) of $2-3$~mV smaller \citep{GaspariniMiglioreMagee_JN2004,Metz_JN2005,Routh_etal_JNeurophys2009}. The peak of action potentials is artificially set in simulations to $v_p=30~mV$ \citep{McDermott_etal_JN2003,Magee_JN1998}, whereas the refractory period for neuronal action potential generation is fixed at $\tau_r=2$~ms \citep{McDermott_etal_JN2003,GaspariniMiglioreMagee_JN2004,Rauch_JNP2003}.
\subsection{Postsynaptic currents}
AMPA~receptor-mediated EPSCs recorded at central synapses are characterized by small rise time constants ($\tau_N^r$), namely in between~0.2--0.6~ms \citep{Spruston_JP1995,MageeCook_NatureNeurosci2000,AndrasfalvyMagee_JN2001,McDermott_etal_JP2006}, and decay time constants ($\tau_N$) in the range of~2.7--11.6~ms \citep{AndrasfalvyMagee_JN2001,PankratovKrishtal_JN2003,Smith_etal_JP2003,McDermott_etal_JP2006}. Furthermore, whole-cell recordings of EPSCs for quantal-like glutamatergic stimulation report amplitudes for these currents generally within $\sim$15.5--30~pA \citep{PankratovKrishtal_JN2003,McDermott_etal_JP2006}, and corresponding somatic depolarizations (EPSPs) similarly are in a wide range of values comprised between $\sim$0.5--7.2~mV, consistent with large quantal size variability of glutamate release from presynaptic terminals \citep{Loebel_FCN2009}. Accordingly, in the simulations of \figref{fig:sics} we set $\tau_N^r=0.5$~ms and $\tau_N=10$~ms and take the two scaling factors $\hat{J}_S,\,\hat{I}_S$ in \eqref{eq:psc} such that \citep{Abbott2002}
\begin{align}
\hat{J}_S &= \frac{J_S}{\varrho_c\, Y_T\, \tau_N}\label{eq:psc-normalization}\\
\hat{I}_S &= I_S \rpar{\frac{1}{\tau_N}-\frac{1}{\tau_N^r}}\rpar{\rpar{\frac{\tau_N^r}{\tau_N}}^{\rpar{\frac{\tau_N}{\tau_N-\tau_N^r}}} - \rpar{\frac{\tau_N^r}{\tau_N}}^{\rpar{\frac{\tau_N^r}{\tau_N-\tau_N^r}}}}^{-1}
\end{align}
In this fashion, with~3/4 of released neurotransmitter reaching postsynaptic receptors (i.e. $\zeta=0.75$), setting $J_S=4.27$ results in EPSP amplitudes approximately equal to $I_S$. In order to convert PSCs and SICs from voltage to current units, we divide them by typical neuronal input resistance ($R_{in}$) values which are generally reported in the range of $\sim$60--150~M$\Omega$ \citep{Magee_JN1998,McDermott_etal_JP2006,Routh_etal_JNeurophys2009}.   
\subsection{Slow-inward currents}
Astrocyte-mediated SICs are documented in a wide range of amplitudes that spans from $>$10~pA \citep{FellinCarmignoto_Neuron2004,PereaAraque_JN2005,Chen_PNAS2012,Perea_NatComm2014,Martin_Science2015} to $>$200~pA \citep{FellinCarmignoto_Neuron2004,KangKang2005,Bardoni_JP2010,Nie_JNP2010}, although their majority is mostly found between~30--80~pA in physiological conditions \citep{FellinCarmignoto_Neuron2004,Martin_Science2015}. SICs kinetics also likely varies depending on subunit composition of SIC-mediating NMDA receptors \citep{Traynelis_etal_PharmRev2010}. In general, assuming NR2B-containing NMDARs as the main receptor type mediating SICs \citep{Papouin_PTRSB2014}, mean SICs rise and decay times are respectively reported in $\sim$30--90~ms and $\sim$100--800 \citep{Angulo2004,FellinCarmignoto_Neuron2004}. Accordingly, in the simulations of \figref{fig:sics} we set $\tau_S^r=30$~ms and $\tau_N=600$~ms and take the two scaling factors $\hat{J}_S,\,\hat{I}_S$ in \eqref{eq:sic} such that
\begin{align}
\hat{J}_A &= \frac{J_A}{\varrho_e\, G_T\, \tau_S}\\
\hat{I}_A &= I_A \rpar{\frac{1}{\tau_S}-\frac{1}{\tau_S^r}}\rpar{\rpar{\frac{\tau_S^r}{\tau_S}}^{\rpar{\frac{\tau_N}{\tau_N-\tau_S^r}}} - \rpar{\frac{\tau_S^r}{\tau_S}}^{\rpar{\frac{\tau_S^r}{\tau_S-\tau_S^r}}}}^{-1}
\label{eq:sic-normalization}
\end{align}
In this fashion, we ultimately choose the value of $J_S$ to account for SIC-mediated depolarizations approximately equal to $I_A$. As SIC amplitudes are generally reported in terms of current rather than voltage units, we estimate somatic depolarizations ensuing from SICs by expressing these latter in terms of typical EPSCs. Thus, for example, for individual EPSCs of 30~pA that generate 2~mV EPSPs, realistic SICs could be regarded on average to be $\sim$1--5-fold these EPSCs and thus contribute to a similar extent to 1--5~times typical EPSPs, that is $\sim$2--10~mV.
\subsection{Spike-timing--dependent plasticity}
We consider the set of parameters for the nonlinear Ca$^{2+}$ model by \citet[Figure~S6]{Graupner_PNAS2012} originally proposed by these authors to qualitatively reproduce the classic STDP curve \citep{BiPoo_JN1998}. In addition, SIC-mediated postsynaptic Ca$^{2+}$ transients are assumed similar to presynaptically-triggered~Ca$^{2+}$ transients but with likely longer rise and decay times. Finally, both presynaptically-mediated and SIC-mediated Ca$^{2+}$ transients are rescaled by equations analogous to~\ref{eq:psc-normalization}--\ref{eq:sic-normalization} in order to obtain transient peak amplitudes equal to $C_{pre}$ and $C_{sic}$ respectively. 
\newpage
\section{Parameter ranges and values}
Values of model parameters used in our simulations are summarized in the following table. Blank table entries are for those parameters whose value was either taken from previously published studies \citep{DePitta_JOBP2009,Graupner_PNAS2012,Wallach_PCB2014} or estimated on the basis of other model parameters (see \appref{app:par-estimation}). Simulation specific~(s.s.) parameter values are instead specified within figure captions.
\begin{longtable}{l l@{  } l l@{ } l@{  }}
    \hline
    Symbol      & Description             &Range       &Value     &Units\\
    \hline \endfirsthead

    \caption[]{\emph{continued}}\\
    \hline
    Symbol      & Description             &Range       &Value     &Units\\
    \hline \endhead

    \hline \multicolumn{4}{r}{\emph{continued on the next page}} \endfoot

    \hline \endlastfoot

    \multicolumn{5}{c}{\textsl{Synaptic dynamics}}\\
    $\tau_{d}$ & Depression time constant                &$>$0.01--2   &s.s. &s\\
    $\tau_{f}$ & Facilitation time constant              &$>$0.5--2    &s.s. &s\\
    $U_{0}$    & Resting synaptic release probability    &$<$0.09--0.9 &s.s. &--\\
    \multicolumn{5}{c}{\textsl{Neurotransmitter release and time course}}\\
    $Y_{T}$      & Total vesicular glutamate concentration &300--1000  &500			 &mM\\
    $\varrho_{c}$& Vescicular vs. mixing volume ratio      & 		   &0.005        &--\\
    $\tau_{c}$   & Glutamate clearance time const.         &2--100     &25			 &ms\\
    $\zeta$      & Efficacy of synaptic transmission       &0--1       &0.75         &--\\ 
    \multicolumn{5}{c}{\textsl{Astrocyte GPCR kinetics}}\\
    $O_{A}$     & Agonist binding rate	                   &    &0.3  &$\mu$M$^{-1}$s$^{-1}$\\
    $\tau_{A}$  & Agonist unbinding time                   &    &0.55 &s\\
    \multicolumn{5}{c}{\textsl{IP$_{3}$R kinetics}}\\
    $O_{2}$     & Inact.~Ca$^{2+}$ binding rate (with Ca$^{2+}$~act.)&0.04--0.18  &0.2 &$\mu$M$^{-1}$s$^{-1}$\\
    $d_{1}$     & IP$_{3}$ binding affinity                          &0.1--0.15   &0.13&$\mu$M\\
    $d_{2}$     & Inact.~Ca$^{2+}$ binding affinity (Ca$^{2+}$~act.) &        &1.05   &$\mu$M\\
    $d_{3}$     & IP$_{3}$ binding affinity~(Ca$^{2+}$~inact.)       &        &0.9434 &$\mu$M\\
    $d_{5}$     & Act.~Ca$^{2+}$ binding affinity                    &        &0.08   &$\mu$M\\
    \multicolumn{5}{c}{\textsl{Calcium fluxes}}\\
    $\varrho_{A}$  & ER-to-cytoplasm volume ratio             &0.4--0.7      &0.18   &--\\
    $C_{T}$     & Total~ER~Ca$^{2+}$ content                  &3--5          &2      &$\mu$M\\
    $\Omega_{L}$& Max.~Ca$^{2+}$ leak rate                    &0.05--0.1     &0.1    &s$^{-1}$\\
    $\Omega_{C}$& Max.~Ca$^{2+}$ release rate by~IP$_{3}$Rs   &$>$6          &6      &s$^{-1}$\\
    $K_{P}$     & Ca$^{2+}$ affinity of SERCA pumps           &0.05--0.1     &0.05   &$\mu$M\\
    $O_{P}$     & Max.~Ca$^{2+}$ uptake rate                  &0.4--1.3      &0.9    &$\mu$M$\,$s$^{-1}$\\
    \multicolumn{5}{c}{\textsl{IP$_{3}$ production}}\\
    $O_{\beta}$ & Max. rate of~IP$_{3}$ production by~PLC$\beta$ &0.05--2    &1        &$\mu$M$\,$s$^{-1}$\\
    $K_{\delta}$&~Ca$^{2+}$ affinity of~PLC$\delta$              &0.1--1     &0.5      &$\mu$M\\
    $\kappa_{\delta}$ & Inhibiting~IP$_{3}$ affinity of~PLC$\delta$  &1--1.5 &1        &$\mu$M\\
    $O_{\delta}$      & Max. rate of~IP$_{3}$ production by~PLC$\delta$ &$<$0.8  &0.05 &$\mu$M$\,$s$^{-1}$\\
    \multicolumn{5}{c}{\textsl{IP$_{3}$ degradation}}\\
    $\Omega_{5P}$     & Max. rate of~IP$_{3}$ degradation by~IP-5P&$>$0.05--0.25 &0.1&s$^{-1}$\\
    $K_{D}$           &~Ca$^{2+}$ affinity of~IP$_{3}$-3K         &0.4--0.5      &0.5&$\mu$M\\
    $K_{3K}$          &~IP$_{3}$ affinity of~IP$_{3}$-3K          &0.7--1        &1  &$\mu$M\\
    $O_{3K}$          & Max. rate of~IP$_{3}$ degradation by~IP$_{3}$-3K &$>$0.6 &4.5&$\mu$M$\,$s$^{-1}$\\
    \multicolumn{5}{c}{\emph{Gliotransmitter release and time course}}\\
    $C_{\theta}$    & Ca$^{2+}$~threshold for exocytosis     &0.15--0.8     &0.5  &$\mu$M\\
    $\tau_{G}$      & Glutamate recycling time const.        &0.003--1.5    &1.66 &s\\
    $U_{A}$         & Resting glutamate release probability  &$<$0.9        &0.6  &--\\
    $\varrho_{e}$   & Vescicular vs. mixing volume ratio     &              &6.5$\,\cdot 10^{-4}$   &--\\
    $\tau_{e}$      & Glutamate clearance time const.        &$\le$300      &200  &ms\\
    \multicolumn{5}{c}{\emph{Presynaptic receptors}}\\
    $O_{P}$         & Activation rate                   &$>$0.3   &1.5      &$\mu$M$^{-1}$s$^{-1}$\\
    $\tau_{P}$      & Inactivation time const.          &$>$30--180  &120   &s\\
    $\xi$           & Gliotransmission type             &0--1     &s.s.     &--\\
    \multicolumn{5}{c}{\emph{Postsynaptic neuron}}\\
    $\tau_m$        & Membrane time constant            &20--70   &40        &ms\\
    $\tau_r$        & Refractory period                 &1--5     &2         &ms\\
    $E_L$           & Resting potential                 &-78.2$-$-54.8 &-60  &mV\\
    $v_\theta$      & Firing threshold                  &-55$-$-51     &-55  &mV\\
    $v_r$           & Reset potential                   &-58$-$-53     &-57  &mV\\
    $v_p$           & Peak AP amplitude                 &29.8--41.2    &30   &mV\\
    $R_{in}$        & Input resistance				    &60--150	   &s.s. &M$\Omega$\\	
    \multicolumn{5}{c}{\emph{Postsynaptic currents}}\\
    $\tau_N^r$      & EPSC rise time                    &0.4--0.6  &0.5     &ms\\
    $\tau_N$        & EPSC decay time                   &2.7--11.6 &10      &ms\\
    $J_S$           & Synaptic efficacy                 &          &4.3     &--\\ 
	$I_S$           & EPSP amplitude                    &0.5--7.5  &2       &mV\\
    \multicolumn{5}{c}{\emph{Slow inward currents}}\\
    $\tau_S^r$      & SIC rise time                    &20--70   &20        &ms\\
    $\tau_S$        & SIC decay time                   &100--800 &600       &ms\\
	$J_A$           & SIC efficacy                     &         &68        &--\\
	$I_A$           & SIC amplitude                    &1--10    &4.5       &mV\\
    \multicolumn{5}{c}{\emph{Spike-timing dependent plasticity}}\\
    $C_{pre}$       & NMDAR-mediated~Ca$^{2+}$ increase per AP   &   &1.0        &--\\
    $\tau_{pre}^r$  & NMDAR~Ca$^{2+}$ rise time                  &   &10         &ms\\
    $\tau_{pre}$    & NMDAR~Ca$^{2+}$ decay time                 &   &30         &ms\\
    $W_N$           & Synaptic weight                            &   &39.7       &--\\
    $C_{post}$      & VDCC-mediated~Ca$^{2+}$ increase per AP    &   &2.5        &--\\
    $\tau_{post}^r$ & VDCC~Ca$^{2+}$ rise time                   &   &2          &ms\\
    $\tau_{post}$   & VDCC~Ca$^{2+}$ decay time                  &   &12         &ms\\
    $C_{sic}$       & SIC-mediated~Ca$^{2+}$ increase per AP     &   &1.0        &--\\
    $\tau_{sic}^r$  & SIC~Ca$^{2+}$ rise time                    &   &5          &ms\\
    $\tau_{sic}$    & SIC~Ca$^{2+}$ decay time                   &   &100        &ms\\
    $W_A$           & SIC weight                            &   &10.6           &--\\
    $\eta$          & Amplification of NMDAR-mediated~Ca$^{2+}$  &   &4          &--\\
    $\theta_d$      & LTD threshold                         &   &1.0        &--\\
    $\theta_p$      & LTP threshold                         &   &2.2        &--\\
    $\gamma_d$      & LTD learning rate                     &   &0.57       &s$^{-1}$\\
    $\gamma_p$      & LTP learning rate                     &   &2.32       &s$^{-1}$\\
    $\rho_\star$    & Boundary between UP/DOWN states       &   &0.5        &--\\
    $\tau_\rho$     & Decay time of synaptic change         &   &1.5        &s\\
    $\sigma$        & Noise amplitude                       &   &0.1        &--\\
    $\beta$         & Fraction of synapses in the DOWN state&   &0.5        &--\\
    $b$             & UP/DOWN Synaptic strength ratio       &   &4          &--\\
\end{longtable}
\label{TA:All-Parameters}

\end{appendices}

\newpage
\section{Figure Captions}
\begin{figure}[!tp]
\centering
\includegraphics[width=\textwidth]{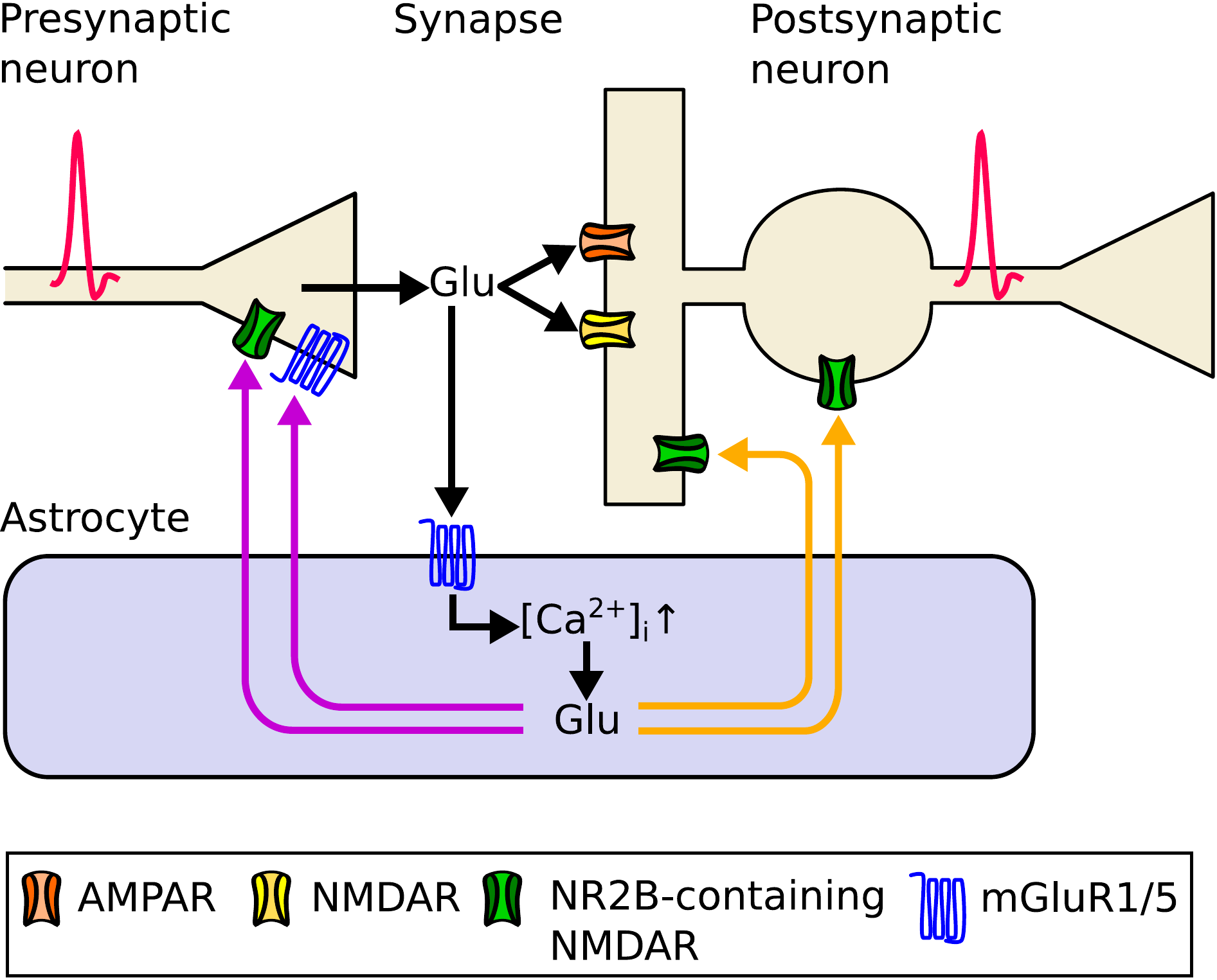}
\caption{Pathways of glutamatergic gliotransmission.}\label{fig:signalling-pathway}
\end{figure}

\textbf{\figref{fig:signalling-pathway}}. Pathways of glutamatergic gliotransmission. Perisynaptic astrocytic processes in several brain areas and different excitatory (but also inhibitory) synapses, may release glutamate in a~Ca$^{2+}$-dependent fashion. In turn, released astrocytic glutamate, may increase (or decrease) synaptic neurotransmitter release by activating extrasynaptically-located presynaptic receptors (\textit{magenta arrows}), or contribute to postsynaptic neuronal depolarization by binding to extrasynaptic NMDA receptors (\textit{orange arrows}) which mediate slow inward currents (SICs). These receptors often (but not always) contain NR2B subunits and are thus different with respect to postsynaptic NMDARs. Glutamate release by the astrocyte could be triggered either by activity from the same synapses that are regulated by the astrocyte (homosynaptic scenario) or by other synapses that are not directly reached by glutamatergic gliotransmission (heterosynaptic scenario).\\

\begin{figure}[!tp]
\centering
\includegraphics[width=\textwidth]{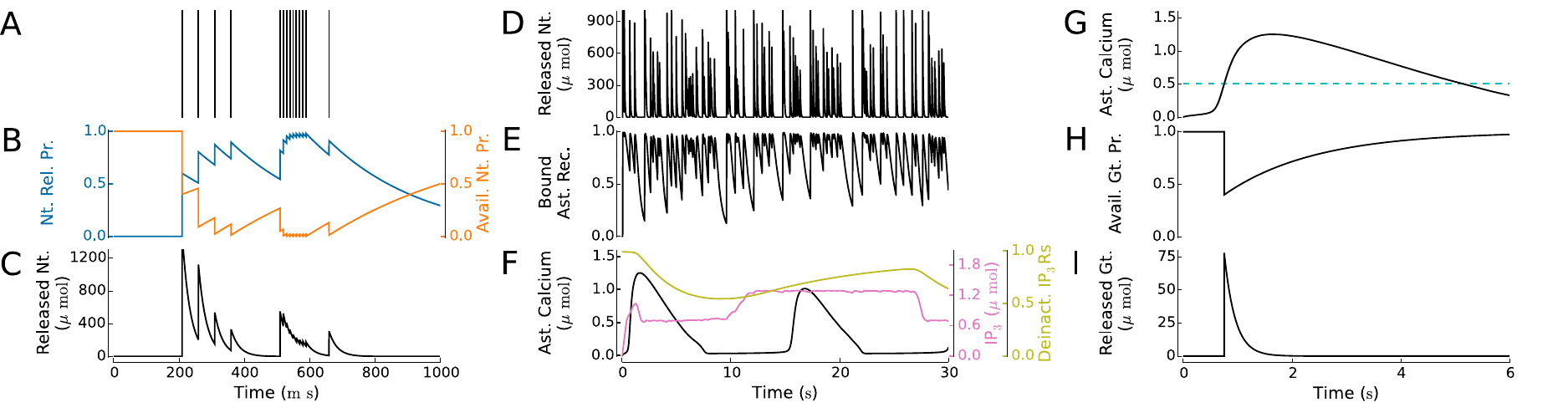}
\caption{Biophysical modeling of a gliotransmitter-regulated synapse.}\label{fig:tripartite_io}
\end{figure}

\noindent
\textbf{\figref{fig:tripartite_io}}. Biophysical modeling of a gliotransmitter-regulated synapse. \textbf{A}-\textbf{C}~Model of synaptic release. Incoming presynaptic spikes (\textbf{A}) increase intrasynaptic Ca$^{2+}$ levels which directly control the probability of release of available neurotransmitter resources (\textbf{B}, Nt.~Rel.~Pr.) and decrease, upon release, the fraction (or probability) of neurotransmitter-containing vesicles available for release (Avail.~Nt.~Pr.). Each spike ensues in release of a quantum of neurotransmitter from the synapse (\textbf{C}, Released~Nt.) whose concentration in the perisynaptic space decays exponentially. Synapse  parameters: $\tau_d=0.5$~s, $\tau_f=0.3$~s, $U_0=0.6$. Stimulation by Poisson-distributed APs with an average rate of~5~Hz. \textbf{D}-\textbf{F}~Model for astrocyte activation. Synaptically-released neurotransmitter in the perisynaptic space (\textbf{D}) binds astrocytic receptors (\textbf{E}, Bound Ast. Rec.), resulting in~IP$_3$ production which triggers~Ca$^{2+}$ signalling in the astrocyte (\textbf{F}). This latter also depends on the fraction of deinactivated~IP$_3$ receptors/Ca$^{2+}$ channels (Deinact.~IP$_3$Rs) on the astrocyte ER membrane (see \appref{app-sec:ca-model}). \textbf{G}-\textbf{I}~Model for gliotransmitter release. The increase of astrocytic~Ca$^{2+}$ beyond a threshold concentration (\textbf{G}, \textit{cyan dashed line}) results in the release of a quantum of gliotransmitter, which decreases the probability of further release of gliotransmitter~(\textbf{H}, Avail.~Gt.~Pr.) while transiently increasing extracellular gliotransmitter concentration (\textbf{I}, Released Gt.). Model parameters as in the Table of \appref{TA:All-Parameters}.\\

\begin{figure}[!tp]
\centering
\includegraphics[width=\textwidth]{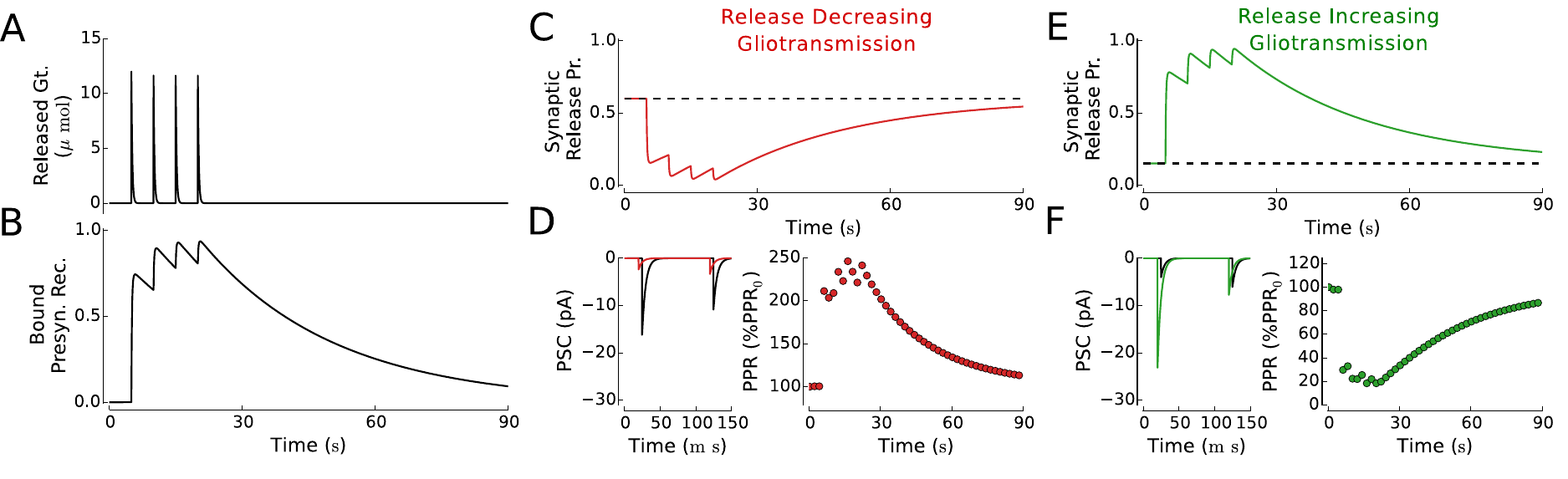}
\caption{Presynaptic pathway of gliotransmission.}\label{fig:pre-loop}
\end{figure}

\noindent
\textbf{\figref{fig:pre-loop}}. Presynaptic pathway of gliotransmission. Gliotransmitter released from the astrocyte (\textbf{A}) binds extrasynaptically-located presynaptic receptors (\textbf{B}) thereby decreasing or increasing synaptic release depending on the type of gliotransmitter and receptor. In the release-decreasing case, synaptic release probability could approach zero by gliotransmission (\textbf{C}, \textit{red trace}, $\xi=0$), although in practice, less dramatic reductions are more likely to be measured with respect to the original value in the absence of gliotransmission (\textit{black dashed line}). The reduction in synaptic release probability, changes pair pulse plasticity increasing the pair pulse ratio (\textbf{D}). In the case of release-increasing gliotransmission, synaptic release probability could instead increase up to one (\textbf{E}, \textit{green trace}, $\xi=1$). In turn, pair pulse plasticity changes towards a decrease of the ensuing pair pulse ratio~(\textbf{F}). Parameters as in \appref{TA:All-Parameters} Table except for $\varrho_e=10^{-4}$, $O_P=0.6$~$\mu$M$^{-1}\,$s$^{-1}$, $\tau_P=30$~s, $\zeta=0.54$,  $J_S=3$~mV, $R_{in}=60$~M$\Omega$.\\

\begin{figure}[!tp]
\centering
\includegraphics[width=\textwidth]{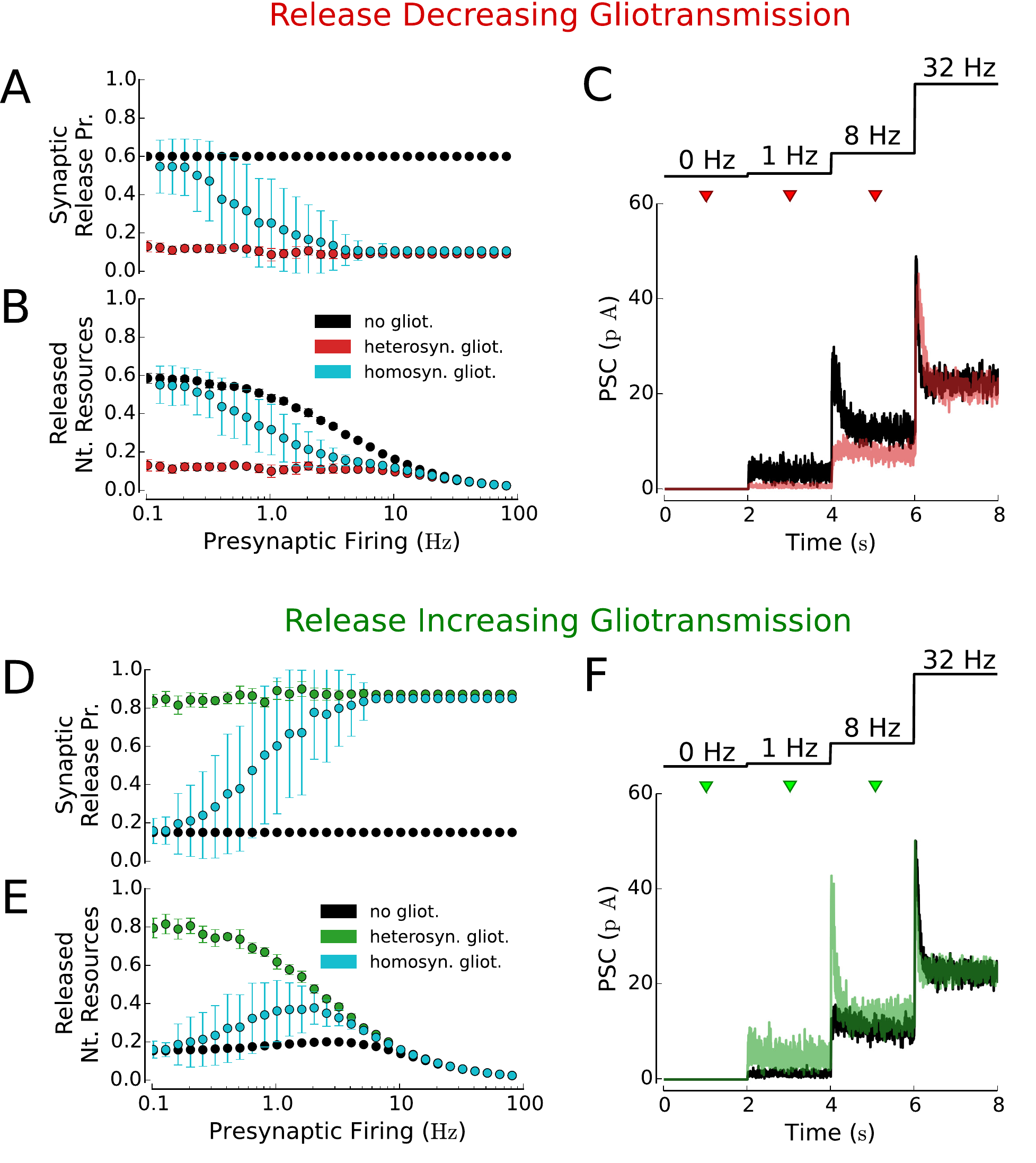}
\caption{Gliotransmitter-mediated modulation of synaptic frequency response.}\label{fig:syn_filtering}
\end{figure}

\noindent
\textbf{\figref{fig:syn_filtering}}. Gliotransmitter-mediated modulation of synaptic frequency response. Decrease (\textbf{A}) or  increase (\textbf{D}) of synaptic release probability by gliotransmission modulate the average per-spike synaptic release, resulting in a change of the synapse frequency response. Monotonically-decreasing frequency responses, that are typical of depressing synapses could be flattened by release-decreasing gliotransmission (\textbf{B}, \textit{black} vs.~\textit{red points}), and vice-versa, almost non-monotonic ones, characteristic of facilitating synapses, could turn into monotonically-decreasing responses by release-increasing gliotransmission (\textbf{E}, \textit{black} vs.~\textit{green points}). Changes in frequency response depend on whether gliotransmission impinges on the very synapse that is triggered by (homosynaptic/closed-loop scenario) or not (heterosynaptic/open-loop scenario). In the homosynaptic scenario, the synaptic response is expected to change only for presynaptic firing rates that are sufficiently high to trigger gliotransmitter release from the astrocyte (\textbf{B},~\textbf{E},~\textit{cyan points}). Data points and error bars:  mean$\,\pm\,$STD for $n=20$ (no gliot. and heterosyn. gliot.) or $n=200$ simulations (homosyn. gliot.) with 60~s-long Poisson-distributed presynaptic spike trains. \textbf{C},~\textbf{F}~The change of synaptic frequency response mediated by gliotransmission (three consecutive gliotransmitter releases at the time instants marked by \textit{triangles}) leads to changes in how presynaptic firing rates (\textit{top panels}) are transmitted by the synapse (\textit{bottom panels}). Simulated postsynaptic currents (PSC) are shown as average traces of $n=1000$ simulations for gliotransmitter release at 1~Hz. Release-decreasing gliotransmission was achieved for $\xi=0$, whereas $\xi=1$ was used for release-increasing gliotransmission. Depressing synapse in \textbf{A},~\textbf{B}: $\tau_d=0.5$~s, $\tau_f=0.3$~s, $U_0=0.6$; facilitating synapse in \textbf{D},~\textbf{E}: $\tau_d=0.5$~s, $\tau_f=0.5$~s, $U_0=0.15$. Other model parameters as in \figref{fig:pre-loop} except for $R_{in}=300$~M$\Omega$.\\

\begin{figure}[!tp]
\centering
\includegraphics[width=\textwidth]{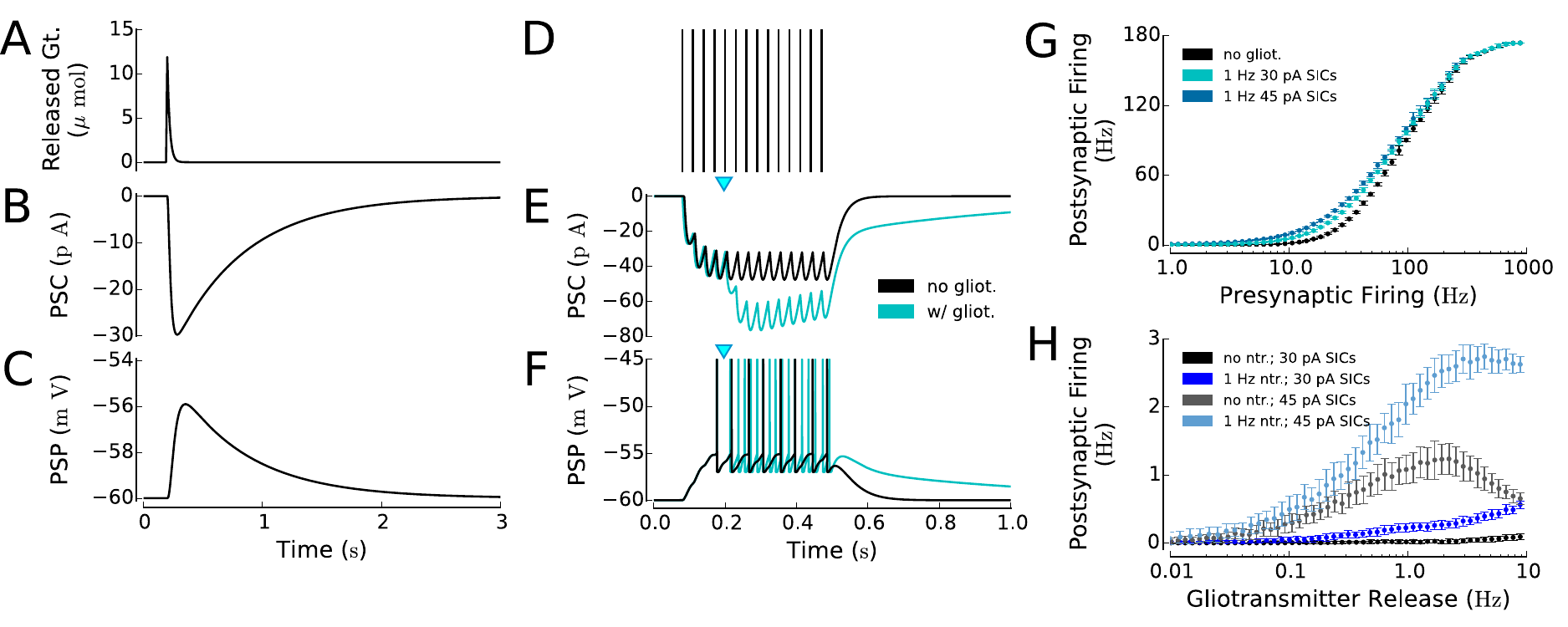}
\caption{Postsynaptic pathway of gliotransmission by slow inward currents.}\label{fig:sics}
\end{figure}

\noindent
\textbf{\figref{fig:sics}}. Postsynaptic pathway of gliotransmission by slow inward currents. The transient increase of gliotransmitter concentration in the perisynaptic space (\textbf{A}), triggers a slow inward (depolarizing) current (SIC) in the postsynaptic neuron (\textbf{B},~\textbf{C}). Such~SIC adds to postsynaptic currents triggered by presynaptic spikes (\textbf{D},~\textbf{E}, \textit{cyan triangle} marks gliotransmitter release/SIC onset) and may dramatically alter postsynaptic firing (\textbf{F}). In general postsynaptic firing frequency increases both with~SIC amplitude (\textbf{G}) and frequency (\textbf{H}). In this latter case however, SICs as ample as 30~pA (similar to what reported in several experiments) need to impinge on the postsynaptic neuron at unrealistically high rates ($\gg$0.1~Hz) in order to trigger a sensible change in the neuron's firing rate (\textit{black data points}). Lower, more realistic SIC rates may affect neuronal firing only for larger SIC amplitudes (e.g.~45~pA, \textit{grey data points}). The entity of SIC-mediated increase of postsynaptic neuronal firing further depends on the neuron's state of depolarization at SIC timings which is set by synaptic inputs (\textit{blue} and \textit{cyan data points}). Data points and error bars: mean$\,\pm\,$STD out of $n=50$ simulations with presynaptic Poisson-distributed spike trains. Parameters as in \appref{TA:All-Parameters} Table except for $\varrho_e=10^{-4}$, $\tau_e=200$~ms, $\tau_S^r=10$~ms, $R_{in}=150$~M$\Omega$.\\

\begin{figure}[!tp]
\centering
\includegraphics[width=\textwidth]{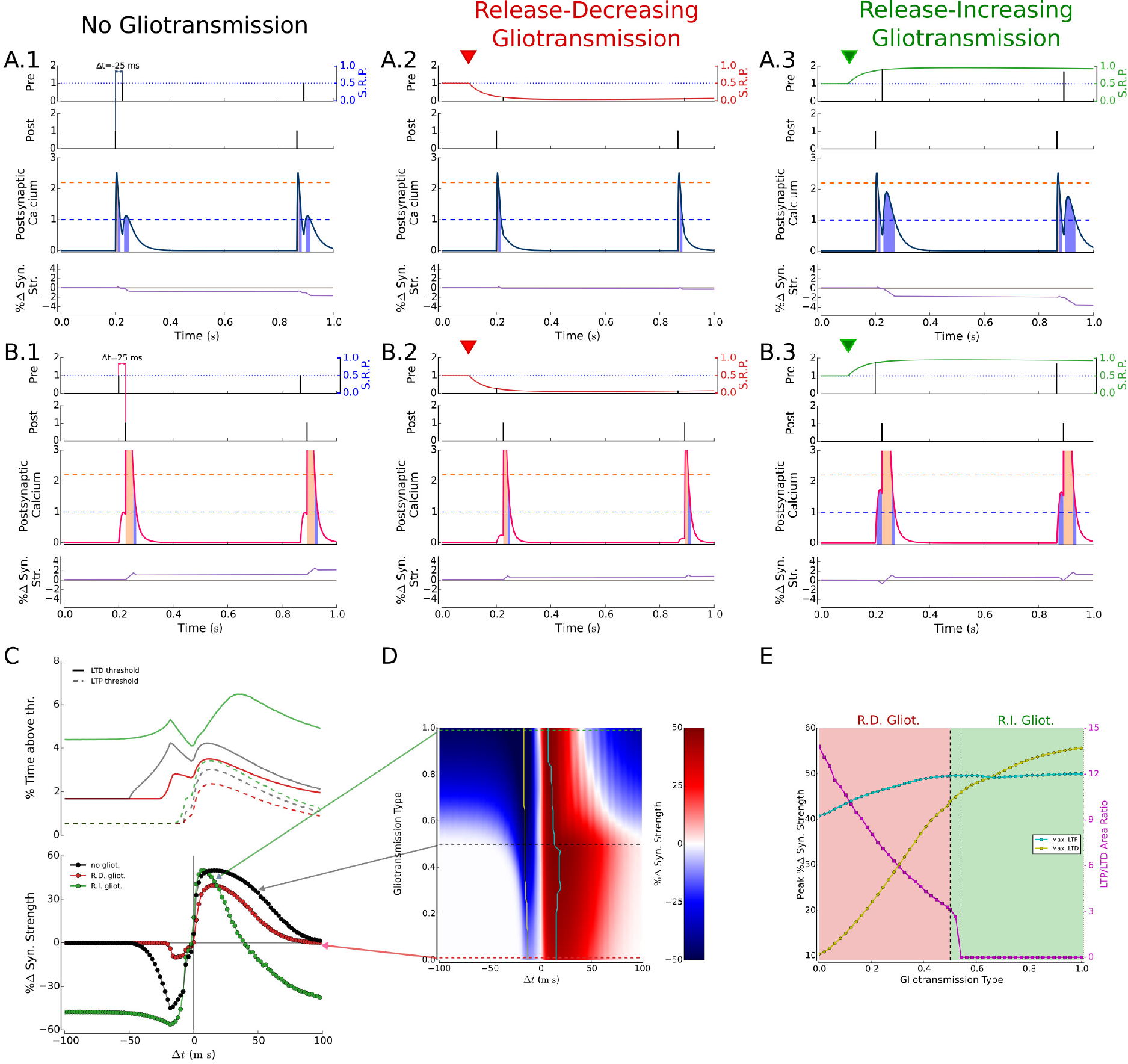}
\caption{STDP Modulation by gliotransmitter regulation of synaptic release.}\label{fig:stdp-curves}
\end{figure}

\noindent
\textbf{\figref{fig:stdp-curves}}.~STDP Modulation by gliotransmitter regulation of synaptic release. (\textbf{A},~\textbf{B})~Rationale of~LTD and~LTP without (\textbf{A.1}, \textbf{B.1}) and with either release-decreasing~(\textbf{A.2},~\textbf{B.2}, $\xi=0$) or release-increasing gliotransmission (\textbf{A.3}, \textbf{B.3}, $\xi=1$) setting on at the \textit{red}/\textit{green marks}. \textbf{C}~Percentage of time spent by postsynaptic Ca$^{2+}$ transients (\textit{top panel}) above depression (\textit{dashed lines}) and potentiation thresholds (\textit{solid lines}) for spike timing intervals ($\Delta t$) between $\pm\,$100~ms, and resulting STDP curves (\textit{bottom panel}) in the absence of gliotransmission (no gliot., \textit{black curve}) and with maximal release-decreasing (R.D., \textit{red curve}) or release-increasing gliotransmission (R.I., \textit{green circles}). \textbf{D}~In general, strength and direction (i.e.~``type'') of gliotransmission may dramatically modulate STDP. For example, synaptic changes are attenuated when synaptic release is decreased by gliotransmission (area below the \textit{black dashed line}). Conversely, for sufficiently strong release-increasing gliotransmission (area above the \textit{black dashed line}), the LTP window shrinks and LTD may be measured for all $\Delta t <0$, as well as for sufficiently large $\Delta t>0$. \textbf{E}~A closer inspection of STDP curves indeed reveals that LTD (\textit{yellow curve}) increases for larger synaptic release accounted by gliotransmission, while the ratio between areas underneath the LTP and LTD (\textit{magenta curve}), initially in favor of the former (i.e. for release-decreasing gliotransmission), reduces to zero for large enough release-increasing gliotransmission, when two open LTD windows appear outside a small LTP window center for small $\Delta t >0$ (\textit{hatched area}). Synaptic parameters: $\tau_d=0.33$~s, $\tau_f=0.33$~s, $U=0.5$~s. Other parameters as in \appref{TA:All-Parameters} Table except for $\varrho_e=10^{-4}$, $\tau_c=1$~ms, $W_N=78.7$, $\tau_P=5$~s in~\textbf{A},~\textbf{B} and $\tau_P=30$~s otherwise.\\

\begin{figure}[!tp]
\centering
\includegraphics[width=.7\textwidth]{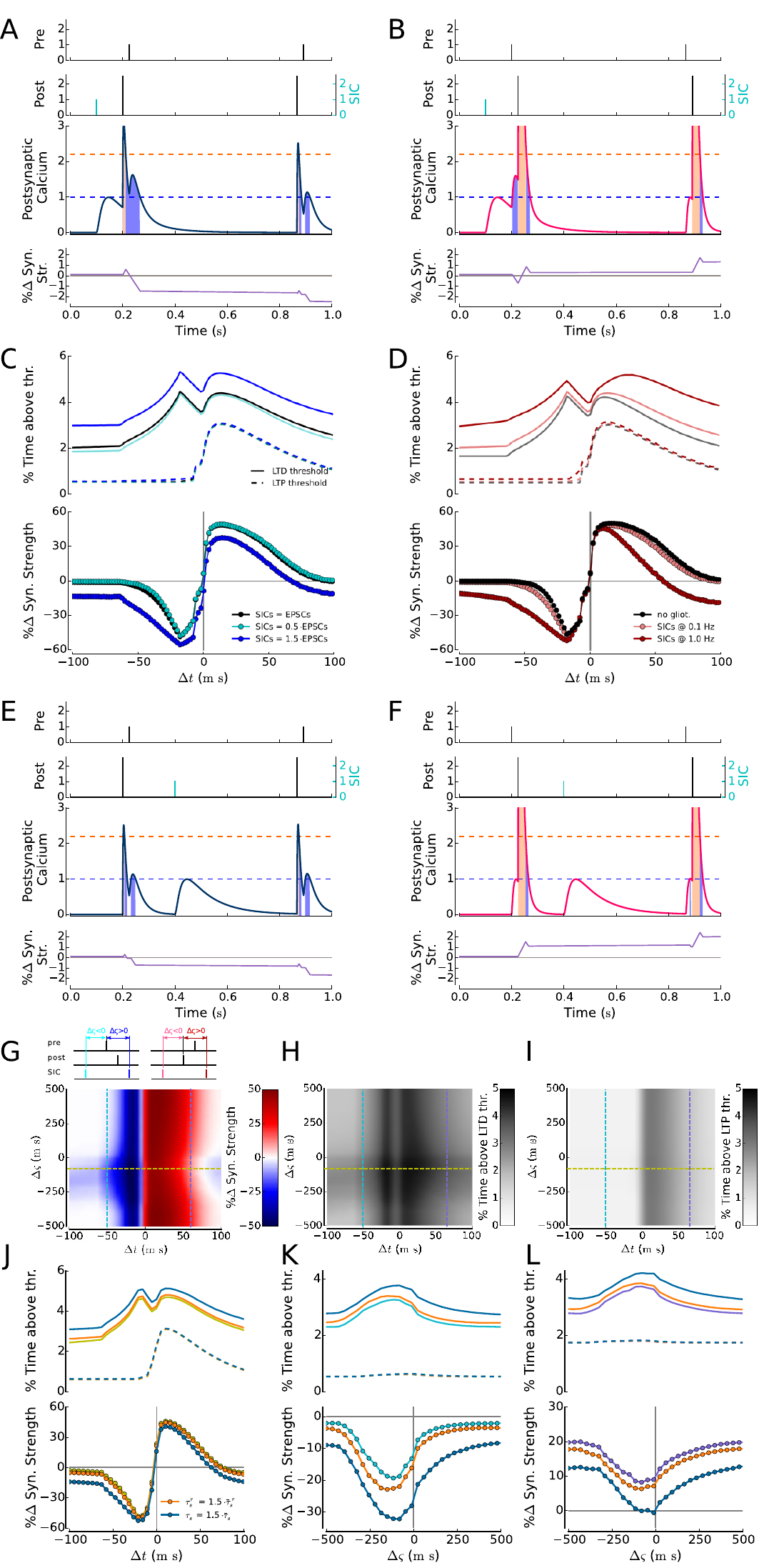}
\caption{STDP modulation by gliotransmitter-mediated~SICs.}\label{fig:stdp_sic}
\end{figure}

\noindent
\textbf{\figref{fig:stdp_sic}}.~STDP modulation by gliotransmitter-mediated~SICs. \textbf{A},~\textbf{B} Inspection of postsynaptic Ca$^{2+}$ in the initial part of a pairing protocol that includes a gliotransmitter-mediated slow inward current (SIC) arriving to the postsynaptic neuron at $t=0.1$~s, illustrates how SICs have the potential to modulate postsynaptic~Ca$^{2+}$ thereby regulating~LTD and LTP. \textbf{C}~The magnitude of modulation depends on how large~SICs are with respect to synaptic inputs (EPSCs) as well as at \textbf{D}~what rate they occur. \textbf{E},~\textbf{F} Impact of the delay ($\Delta \varsigma$) at which~SICs occur with respect to pre/post pairs. \textbf{G}~STDP curves as a function of the~SIC-pre/post pair delay ($\Delta \varsigma$) show how~LTD could get stronger while the~LTP window shrink for small-to-intermediate $\Delta \varsigma \le 0$ in correspondence with \textbf{H},~\textbf{I} a maximum of the duration of~Ca$^{2+}$ transients above the~LTD threshold. These results were obtained assuming~SIC rise and decay time constants respectively equal to $\tau_s^r=\bar{\tau}_s^r=20$~ms and $\tau_s=\bar{\tau}_s=200$~ms. \textbf{J}-\textbf{L}~Peak and range of this~LTD increase ultimately depend on~SIC kinetics as reflected by the change of sample curves for specific $\Delta\varsigma$ (\textit{yellow curve}) and spike timing intervals (\textit{cyan} and \textit{purple curves}) when~SIC rise and/or decay time constants was slowed down~1.5-fold (\textit{orange} and \textit{blue curves} respectively). \textbf{C},~\textbf{D}~STDP curves were calculated for~60 pre/post pairings at~1~Hz and included SICs starting~0.1~s before the first pairing and occurring at~0.1~Hz. The same pairing protocol but with SIC frequency of~0.2~Hz was used instead in figures \textbf{G}--\textbf{L} although SIC onset and kinetics were varied respectively according to $\Delta \varsigma,\,\tau_S^r$ and~$\tau_S$. Synaptic parameters: $\tau_d=0.33$~s, $\tau_f=0.33$~s, $U_0=0.5$~s. Other parameters as in \appref{TA:All-Parameters} Table except for $\varrho_e=10^{-4}$, $\tau_c=1$~ms, $\tau_e=200$~ms, $\tau_S^r=5$~ms, $\tau_S=100$~ms.

\newpage
\clearpage

\newpage
\clearpage

\newpage
\clearpage

\newpage
\clearpage

\newpage
\clearpage

\newpage
\clearpage

\newpage
\clearpage

\bibliography{../myBib.bib}
\end{document}